\documentclass[aps,prc,twocolumn,amsmath,floatfix,nofootinbib]{revtex4-2}

\usepackage{graphicx}
\usepackage[urlcolor=green]{hyperref}

\begin{document}

\author{Hendrik Roch} \email{hroch@physik.uni-bielefeld.de}
\author{Nicolas Borghini} \email{borghini@physik.uni-bielefeld.de}
\affiliation{Fakult\"at f\"ur Physik, Universit\"at Bielefeld, 
	Postfach 100131, D-33501 Bielefeld, Germany}

\title{Fluctuations of anisotropic flow from the finite number of rescatterings in a two-dimensional massless transport model}

\begin{abstract}
We investigate the fluctuations of anisotropic transverse flow due to the finite number of scatterings in a two-dimensional system of massless particles.
Using a set of initial geometries from a Monte Carlo Glauber model, we study how flow coefficients fluctuate about their mean value at the corresponding eccentricity, for several values of the scattering cross section. 
We also show how the distributions of the second and third event planes of anisotropic flow about the corresponding participant plane in the initial geometry evolve as a function of the mean number of scatterings in the system.
\end{abstract}

\maketitle

\section{Introduction}
\label{s:intro}

Anisotropic flow is one of the most prominent observables in heavy-ion collision experiments, especially at high energies. 
The signal, a modulation in the transverse emission pattern of particles, is believed to be largely due to the lack of cylindrical symmetry of the initial geometry, which is converted by rescatterings into an anisotropy in momentum space~\cite{Heinz:2013th,Luzum:2013yya,Bhalerao:2020ulk}.

When the two nuclei collide, the energy they deposit in the region where they overlap is not evenly distributed. 
The random positions at the instant of the collision of the nucleons, and of the underlying quarks and gluons with their associated color fields, result in an inhomogeneous energy-density profile in the overlap zone. 
Restricting oneself to the transverse geometry of that zone, described by polar coordinates $(r,\theta)$, its azimuthal asymmetries are usually characterized by eccentricities~\cite{Alver:2010gr,Teaney:2010vd,Gardim:2011xv}
\begin{equation}
\varepsilon_n {\rm e}^{{\rm i}n\Phi_n} \equiv 
-\frac{\langle r^n{\rm e}^{{\rm i}n\theta}\rangle}{\langle r^n\rangle},
	\label{eq:eccentricities}
\end{equation}
where the angular brackets denote an average over the centered energy (or sometimes entropy) density, while the definition given here holds for $n\geq 2$.
$\Phi_n$ is the azimuth of the so-called $n$-th participant plane.

In turn, the anisotropies of the transverse momentum distribution of particles are quantified by the successive coefficients of a Fourier series~\cite{Voloshin:1994mz}
\begin{equation}
v_n {\rm e}^{{\rm i}n\Psi_n} \equiv \langle {\rm e}^{{\rm i}n\varphi_{\bf p}}\rangle,
\label{eq:def_vn}
\end{equation}
with $\Psi_{n}$ the angle of $n$-th ``event plane''.
Here the brackets denote an average over the momentum distribution of particles in one event, with $\varphi_{\bf p}$ the azimuth of a particle momentum.

Model studies either within hydrodynamics (ideal or dissipative) or kinetic transport approaches have shown that elliptic flow $v_2$ and triangular flow $v_3$ are mostly governed by the corresponding eccentricity through a simple linear relation~\cite{Ollitrault:1992bk,Alver:2010gr,Teaney:2010vd,Borghini:2010hy,Gardim:2011xv,Niemi:2012aj,Plumari:2015cfa}
\begin{align}
v_2&\simeq\mathcal{K}_{2,2}\varepsilon_2, \label{eq:v2_vs_eps2_lin}\\
v_3&\simeq\mathcal{K}_{3,3}\varepsilon_3. \label{eq:v3_vs_eps3_lin}
\end{align}
In peripheral collisions, in which $\varepsilon_2$ is largest, a sizable nonlinear contribution from $\varepsilon_2$ appears in $v_2$~\cite{Noronha-Hostler:2015dbi}
\begin{equation}
	v_2\simeq\mathcal{K}_{2,2}\varepsilon_2+\mathcal{K}_{2,222}\varepsilon_2^3.
	\label{eq:v2_vs_eps2_nonlin}
\end{equation}

As stated above, accounting for the irregularities in the initial geometry of a collision is essential to properly estimate the eccentricity in a given harmonic~\cite{Miller:2003kd,Alver:2010gr}. 
This randomness in the initial state is often modeled with the help of Monte Carlo (MC) Glauber models~\cite{Miller:2007ri}, which is the approach we use in this paper. 
In actual measurements of anisotropic flow, the Fourier coefficients are obtained by averaging over many events with varying initial geometries. 
The event-by-event fluctuations of the eccentricities, which can be studied within a given model of the initial state~\cite{Yan:2014nsa}, result in corresponding fluctuations of the flow harmonics within the set of events used for a specific analysis. 
These flow fluctuations are being extensively investigated, both experimentally~\cite{Aad:2013xma,ALICE:2016kpq,Sirunyan:2017fts,Aaboud:2019sma} and theoretically~\cite{Petersen:2010cw,Qiu:2011iv,Gale:2012rq,Niemi:2012aj,Teaney:2013dta,Ma:2014xfa,Niemi:2015qia,Plumari:2015cfa,Noronha-Hostler:2015dbi,Giacalone:2016eyu,Bhalerao:2018anl}.

Most of the phenomenological studies of fluctuations employ hydrodynamics to describe the main stage of the system dynamics, which reflects the overall successful description of bulk observables --- especially anisotropic flow --- in hydrodynamic studies~\cite{Heinz:2013th}.
However, one could argue that hydrodynamics is possibly too deterministic for a study of fluctuations. 
Assuming the existence of an underlying particle-based description of the system, hydrodynamics corresponds to the regime in which the (quasi)particles undergo very many scatterings. 
Thus, hydrodynamics\footnote{Here we refer implicitly to ``deterministic'' hydrodynamics. How ``fluctuating hydrodynamics''~\cite{Kapusta:2011gt} encodes in its fluctuating fields the fluctuations of an underlying microscopic theory is unclear to us.} does not capture the possible effect of having only a finite number of collisions per particle --- which could even become quite small for collisions of smaller systems, in which anisotropic flow and its fluctuations are well established~\cite{Nagle:2018nvi}.

In this article, we focus on the fluctuations of the anisotropic flow harmonics that arise due to the finite number of rescatterings $N_{\rm resc.}$ per particle in a kinetic-transport description of the system.
We investigate the influence of the average number of particle scatterings on the relationship between elliptic or triangular flow and the corresponding eccentricity.
That is, we study how the distribution of $v_n$ at a given $\varepsilon_n$ --- i.e.\ the conditional probability distribution $p_{v|\varepsilon}(v_n|\varepsilon_n)$ --- is affected by the average number of rescatterings per particle, extending the exploratory study of Ref.~\cite{Vogel:2007yq}.

In the next section we introduce our approach, which consists of two main components:
the calculation of an energy density profile from a MC Glauber model and its conversion into a system of massless test particles (Sec.~\ref{sec:initial_state}) and a two-dimensional transport model for the subsequent evolution of the system (Sec.~\ref{sec:transport}).
Section~\ref{sec:analysis} describes our approach for the analysis of the conditional probability distribution $p_{v|\varepsilon}(v_n|\varepsilon_n)$ as a function of the average number of rescatterings per particle.
In Sec.~\ref{sec:results} we present the results of our statistical analysis in terms of the moments of the conditional probability distribution. 
We also show how the distributions of event-plane angles relative to the participant-plane angles evolve with the number of rescatterings. 
Additional results are presented in the Appendices.
Eventually we summarize our main findings and discuss them in Sec.~\ref{sec:summary}.

\section{Setup of the simulations}
\label{sec:setup}

In this section we describe the details of the preparation of the initial states we use, and we introduce the transport code that evolves these initial conditions.

\subsection{Initial state}
\label{sec:initial_state}

For the initial states for our transport calculations, we used the TGlauberMC code \cite{Loizides:2014vua} to generate overlap geometries of two lead nuclei. 
Using the cross section $\sigma^{\mathrm{inel}}_{\mathrm{NN}}=(67.6\pm 0.6)$ mb of inelastic nucleon-nucleon collisions at $\sqrt{s}=5.02$ TeV, we simulated three sets of $10^4$ events. 
The impact parameter $b$ was fixed in each set, at the values $b=0$, 6 and 9~fm, respectively, while its direction always lies along the $x$-axis. 

From the nucleon positions $(x,y)$ in the transverse plane and the local numbers of binary collisions $N_{\rm coll.}(x,y)$ and of participants $N_{\rm part.}(x,y)$, we first generated an approximate energy density distribution. 
We used an overall scale factor such that the ``hottest spot'' with the highest energy density corresponds to a temperature of about 800~MeV.
This histogram-like energy density was then smeared as a Gaussian distribution\footnote{The smearing is used to avoid shock fronts in the transport calculation.} with a width of
\begin{equation}
R_\mathrm{N} = \frac{1}{2}\sqrt{\frac{\sigma^{\mathrm{inel}}_{\mathrm{NN}}}{\pi}},
\end{equation}
i.e. approximately 0.7~fm.
Note that $R_\mathrm{N}$ is also the length used in the Glauber model to decide if two nucleons collide.

The next step in the generation of our initial condition is the ''particlization'' of the smooth energy density profile.  
To that effect, we compute the particle-number density $n(x,y)$ corresponding to the energy distribution, using the formula for a two-dimensional ideal gas of massless particles at thermal equilibrium without fugacity.
Particle positions are then sampled from $n(x,y)$ with an acceptance-rejection algorithm.
In each event, we sample $N_\mathrm{p}=5\times 10^5$ test particles, irrespective of the actual total energy in the initial state. 

To implement local energy conservation as accurately as possible while keeping the running time of the transport code manageable, we employed a coarser grid spacing of $0.1$~fm for the particle density grid in the particlization procedure instead of the $0.01$~fm used in the energy density calculation from the Glauber model.
This allows us to minimize the number of cells on the grid with a sizable energy but no test particle, without having to resort to too large values of $N_\mathrm{p}$.

Eventually, we generate the local momentum distribution in the initial state.
The energy content of each cell on the grid is evenly distributed between the test particles in the corresponding cell. 
The particle momenta are rotated by a random angle in the transverse plane to mimic an isotropic initial condition in momentum space.
In fact, due to numerical fluctuations even the overall momentum distribution cannot be exactly isotropic in a single event. 
We ensured the absence of an initial global momentum in each event, i.e.\ the absence of a directed flow $v_1$, but higher-order anisotropies of order $N_\mathrm{p}^{-1/2}$ are unavoidable, which we shall discuss in Sec.~\ref{ss:results_initial}. 

Since we are interested in the dependence on the number of rescatterings in the system of the conditional probability distribution $p_{v|\varepsilon}(v_n|\varepsilon_n)$ describing the anisotropic flow response for a given eccentricity, it would seem preferable to generate initial states with a fixed geometry, instead of our choice. 
We did not follow this approach for two reasons: 
On the one hand, we first tried to use ``artificial'' initial states consisting of distorted Gaussian distributions in the spirit of Ref.~\cite{Borghini:2018xum,Kurkela:2018qeb}, in which the eccentricities are an input. 
However we found out that this led to a non-standard $v_2$ response, namely with a negative cubic term, i.e.\ ${\cal K}_{2,222} < 0$ in Eq.~\eqref{eq:v2_vs_eps2_nonlin}, while realistic initial conditions yield ${\cal K}_{2,222} > 0$~\cite{Noronha-Hostler:2015dbi} (see also Table~\ref{tab:fit-params_b9}).
On the other hand, working with a few ``realistic'' initial conditions, like say a dozen events from the TGlauberMC code, and running the subsequent transport component a large number of times with different seeds for the random-number generator entailed the risk of having picked special initial distributions, that would not reflect the generic behavior. 
Since the overall dependence on the number of rescatterings that we report appears to be consistent over the whole eccentricity range scanned in fixed-impact parameter MC Glauber events, we think that our approach, albeit not optimal, is valid.

\subsection{Two-dimensional massless transport}
\label{sec:transport}

The second main ingredient in our simulation is the transport code that performs the time evolution of the initial condition we have just presented. 
Since the main purpose of our study is to investigate the statistical properties of the physical fluctuations of anisotropic flow arising from the finite number of rescatterings per particle $\langle N_{\rm resc.}\rangle$, we need to simulate a large number of events. 
To keep the total computational time reasonable,\footnote{To be specific, one iteration of the transport algorithm over a single initial condition requires about 213~s CPU time on a 2.2~GHz processor. The parameters chosen for this test are those with the largest number of rescatterings, hence the longest simulation time.} we restricted ourselves to a purely two dimensional system, i.e., our massless test particles live and propagate in the $(x,y)$-plane only.%
\footnote{That is, in the equilibrated regime the particles obey two-dimensional thermodynamics, implying in particular a speed of sound $c_s=1/\sqrt{2}$.}
This obviously means that the absolute flow values that are shown below are not meaningful for phenomenological applications. 
Nonetheless, since anisotropic flow in heavy ion collisions is to a large extent controlled by transverse dynamics, in particular if one considers flow at midrapidity, we are confident that the qualitative trends with  $\langle N_{\rm resc.}\rangle$ that we report hereafter are robust and would persist in fully three-dimensional studies.  

Our two-dimensional code is based on the covariant transport theory algorithm of C.~Gombeaud and J.-Y.~Ollitrault described in Ref.~\cite{Gombeaud:2007ub}, to which we refer for further information, in particular on the smooth approach to the hydrodynamic regime. 
The massless particles we consider are modeled as Lorentz-contracted hard spheres --- actually, since the model is two-dimensional and the particles are massless, they appear as (hard) rods with length $r$~--- that undergo elastic two-to-two collisions.
In the code, the collision times are determined in the laboratory frame and a given scattering is described as a collision between a pointlike particle and a particle of size $2r$.
In two dimensions, the ``hard spheres'' modeling implies that total cross section (with dimension of a length) is $\sigma_{\rm tot}=2r$, while an isotropic differential cross section is ensured by relating the scattering angle $\theta^\star$ in the center-of-momentum frame of a given collision to the impact parameter $d$ by $\theta^\star=\pi(1-d/r)$ --- which also ensures that the two particles do not rescatter at once after a first collision. 
The time step of the evolution is chosen such that the particles can collide only once in a time step.

To preserve covariance and locality in the algorithm, and to remain in the regime in which the system dynamics are described by the Boltzmann kinetic theory with a collision kernel including only two-to-two scatterings, it is important that the system should remain dilute.
This requirement can be quantified by the dimensionless dilution parameter $D$ defined as the ratio of the mean inter-particle distance $n^{-1/2}$ over the mean free path $\ell_\mathrm{mfp}$
\begin{align}
D\equiv\frac{n^{-1/2}}{\ell_\mathrm{mfp}},
\label{eq:dilution}
\end{align}
which has to fulfill $D\ll 1$.
For the actual values of $D$ and of the Knudsen number~\eqref{def_Kn} given in the following, the mean free path is taken as $\ell_\mathrm{mfp}\equiv 1/(n\sigma_{\rm tot})$, using the average density $n$ of the system in the initial condition of the event under consideration.
In all simulations we report hereafter, we found $D\lesssim 0.1$, which according to Ref.~\cite{Gombeaud:2007ub} (see in particular Fig.~2) indeed corresponds to the dilute regime, at least as far as anisotropic flow is concerned.

As already stated, we use $N_\mathrm{p}=5\times 10^5$ test particles per event in our simulations. 
To further decrease the numerical noise on the flow harmonics $v_n$ due to the finite number of particles, every Gaussian-smeared initial energy-density profile was converted into a particle system $N_\mathrm{iter.}=10$ times, and the $N_\mathrm{iter.}$ resulting particle distributions were let to evolve with the transport algorithm independently. 
Running $N_\mathrm{iter.}$ events with the same initial energy-density profile and with $N_{\rm p}$ particles each and averaging the flow values over the $N_\mathrm{iter.}$ iterations is equivalent, from the point of view of the numerical uncertainties on the $v_{n}$, to considering a single event with $N_\mathrm{iter.}\cdot N_\mathrm{p}$ test particles: 
in either case the resulting uncertainty on a $v_n$ is of order
\begin{equation}
	\label{delta_vn}
\delta v_n = \frac{1}{\sqrt{N_\mathrm{iter.}\cdot N_\mathrm{p}}}\simeq 4.5\times 10^{-4}
\end{equation}
(and actually even a factor $\sqrt{2}$ smaller~\cite{Voloshin:1994mz}).
Performing the $N_\mathrm{iter.}$ iterations is however faster in practice, since the running time grows faster than linearly with the particle number $N_{\rm p}$. 

For every impact parameter we simulate $10^4$ MC energy distributions, each of which gives rise to $N_\mathrm{iter.}$ initial conditions with $N_\mathrm{p}$ test particles. 
In turn, each of these initial particle distributions is used 3 times as input to the transport algorithm, namely with 3 different cross sections $\sigma_{\rm tot}$ (or equivalently sizes $r$) for the scatterings of the test particles. 
As we shall see later, the largest cross section leads to a $v_2$ or $v_3$ close to the value in reaches in the hydrodynamic limit, while the other two values correspond to the regime in which $v_2$ and $v_3$ vary significantly with $\sigma_{\rm tot}$. 

Instead of using the cross section, we shall characterize the three types of simulations by the Knudsen number Kn, defined as the ratio of the mean free path $\ell_\mathrm{mfp}$ over a length $R$ characterizing the system size:
\begin{equation}
	\label{def_Kn}
	\mathrm{Kn} \equiv \frac{\ell_\mathrm{mfp}}{R}.
\end{equation}
For the characteristic length $R$, we use the root mean square radius (with energy-density weighting) of the initial overlap region of the two lead nuclei.
That is, $R$ varies event to event, as does the mean free path. 
Accordingly, each value of $\sigma_{\rm tot}$ does not correspond to a single Knudsen number, but to a distribution about a mean value $\langle \mathrm{Kn}\rangle$. 
In practice, the relative fluctuation of Kn about $\langle \mathrm{Kn}\rangle$ is of less than 5\,\%, so we shall ignore it in the following and only quote the mean value  or its inverse. 
Note that the value of $\langle \mathrm{Kn}\rangle$ for a given total cross section $\sigma_{\rm tot}$ does depend on the impact parameter of the lead-lead collision.

\begin{figure}[tb!]
	\includegraphics*[width=\linewidth]{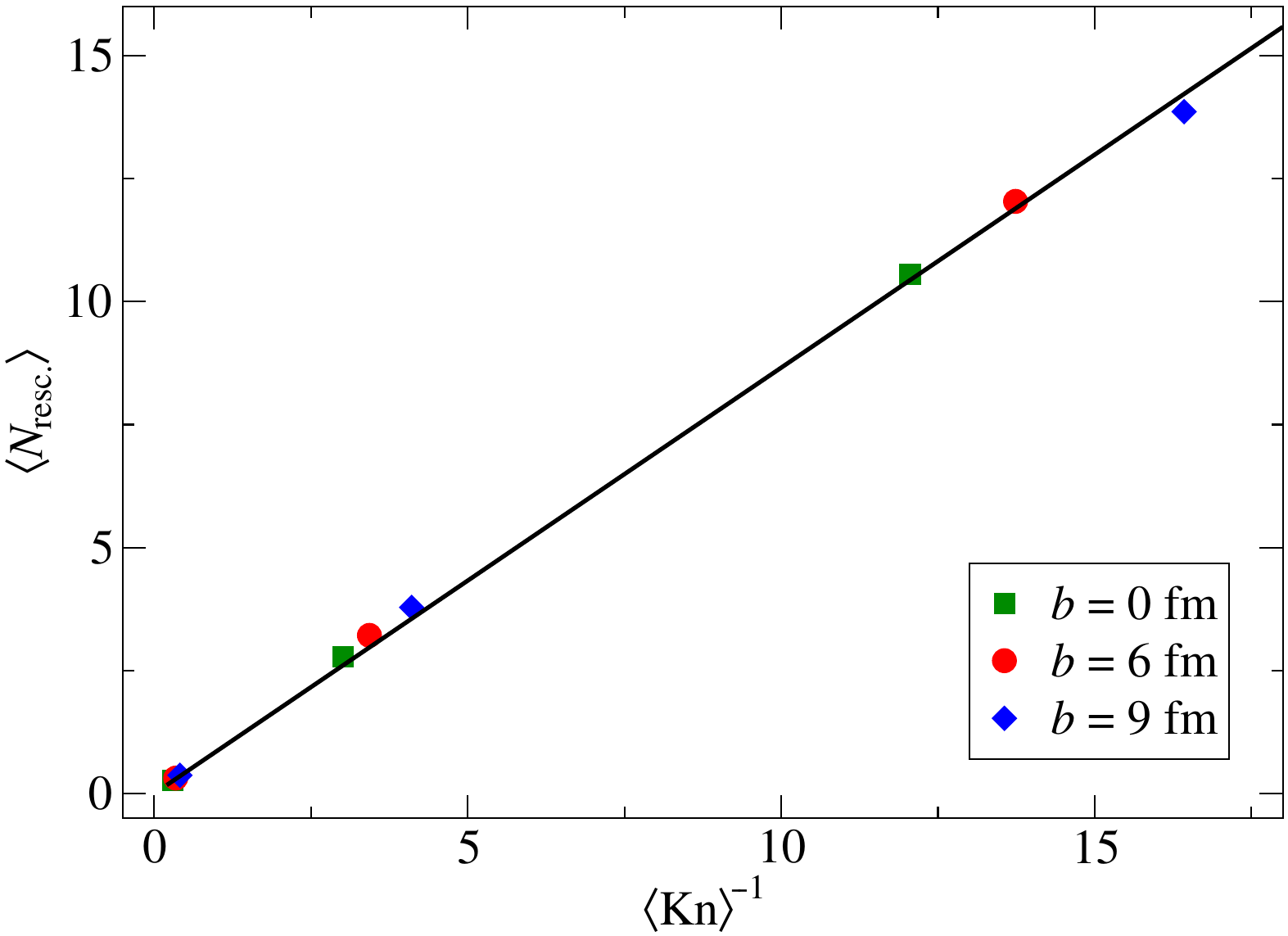}
	
	\caption{Dependence of the average number of rescatterings per particle $\langle N_\textrm{resc.}\rangle$ on the inverse mean Knudsen number $\langle \mathrm{Kn}\rangle^{\!-1}$ for simulations at three different impact parameters and with three different scattering cross sections.
	The full line is the linear fit~\eqref{Nresc_vs_Kn}.}
	\label{fig:Nresc_vs_Kn}
\end{figure}

The Knudsen number, and more precisely its inverse, is related to the average number of rescatterings per particle $\langle N_\textrm{resc.}\rangle$ in the system.
In our simulations, after averaging over events with a fixed impact-parameter value, we find a very simple linear relation
\begin{equation}
\label{Nresc_vs_Kn}
\langle N_\textrm{resc.}\rangle \simeq 0.866\,\langle \mathrm{Kn}\rangle^{\!-1},
\end{equation}
as illustrated in Fig.~\ref{fig:Nresc_vs_Kn}.
Note that $\langle N_\textrm{resc.}\rangle$ depends on the system lifetime. 
In our computations, we let every simulation run for 15~fm/$c$, irrespective of the impact parameter and thus the characteristic length scale $R$.
We checked that using a longer lifetime does not affect our results for the fluctuations of the anisotropic flow coefficients, although it naturally leads to a larger proportionality coefficient in Eq.~\eqref{Nresc_vs_Kn}.

\begin{figure}[!tb]
	\centering
	\includegraphics*[width=\linewidth]{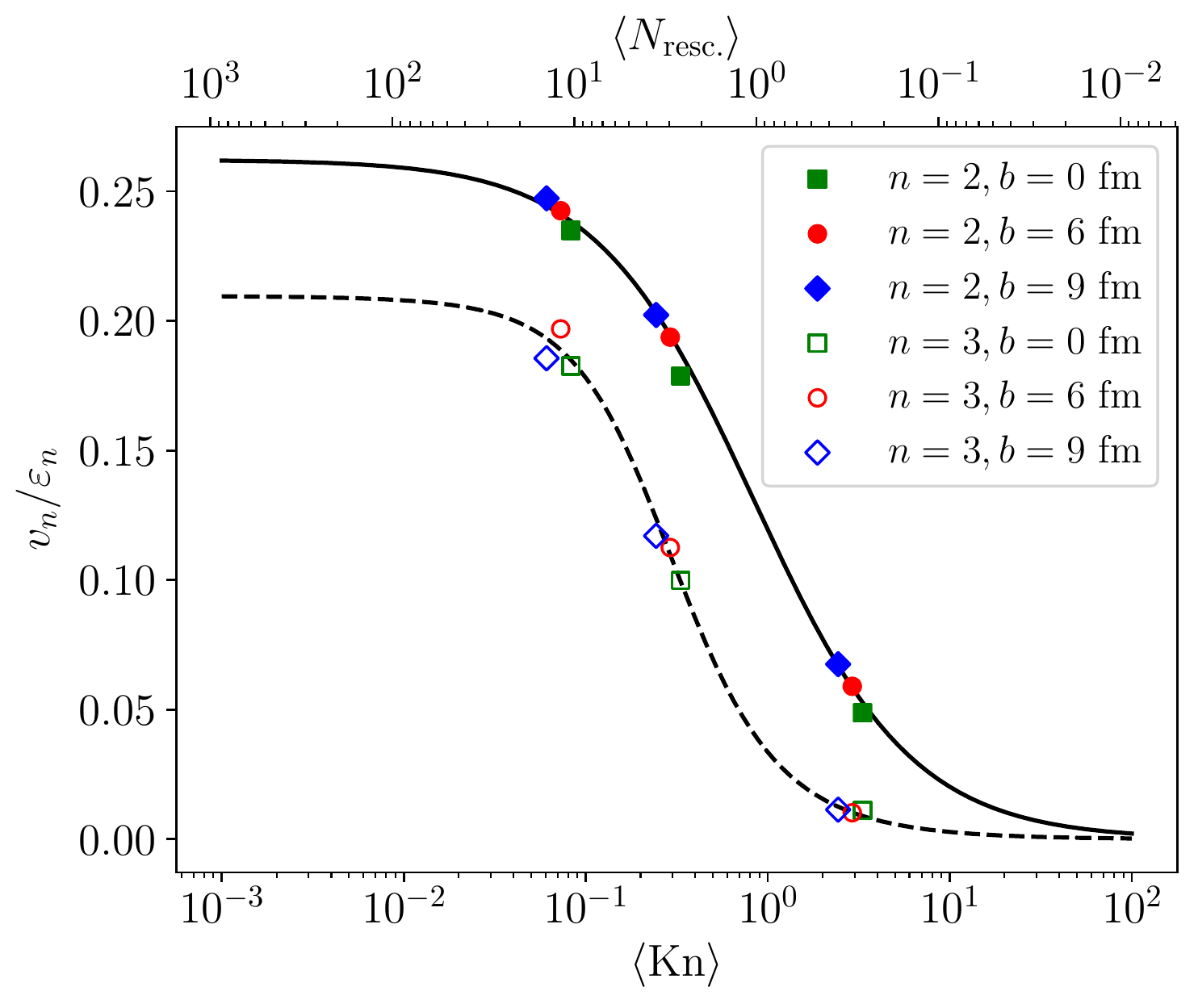}
	\caption{\label{fig:vn_vs_Kn}Elliptic flow $v_2$ and triangular flow $v_3$, divided by the corresponding eccentricity, as a function of the average Knudsen number for simulations with the different cross sections used in this paper. 
	The curves correspond to fits with Eq.~\eqref{eq:v2_vs_Kn} (full curve) resp.\ \eqref{eq:v3_vs_Kn} (dashed curve).}	
\end{figure}

We already mentioned that the three values of the cross section, or equivalently the Knudsen number, that we chose correspond to different regimes for anisotropic flow. 
This is illustrated in Fig.~\ref{fig:vn_vs_Kn}, which shows $v_2$ and $v_3$, divided by the corresponding eccentricity $\varepsilon_2$ or $\varepsilon_3$, as a function of Kn for simulations at the three impact-parameter values. 
On the same plot we display two curves: 
First, a fit through the $v_2/\varepsilon_2$ values with the formula~\cite{Bhalerao:2005mm}
\begin{align}
	v_2 = \frac{v_2^{\mathrm{hydro}}}{1+A_{2\,}\textrm{Kn}},
	\label{eq:v2_vs_Kn}
\end{align}
which was shown~\cite{Gombeaud:2007ub} to provide a good description of a systematical investigation of the behavior of $v_2$ as a function of Kn.%
\footnote{In Ref.~\cite{Gombeaud:2007ub} this was checked using simulations at constant dilution parameter $D$ across the values of Kn. Here our values of $D$ depend on Kn, but stay in the regime in which this does not affect the $v_2$ value.}
Here $v_2^{\rm hydro\!}/\varepsilon_2 = 0.2621$ and $A_2 = 1.1967$.
Secondly, we describe the behavior of $v_3$ as a function of the number of rescatterings with the higher-order Pad\'e approximation~\cite{Alver:2010dn}
\begin{align}
	v_3 = \frac{(1 + B_{3\,}\textrm{Kn})v_3^{\mathrm{hydro}}}{1+(A_3+B_3)\textrm{Kn}+C_{3\,}\textrm{Kn}^2}.
	\label{eq:v3_vs_Kn}
\end{align}
The fit parameters used in Fig.~\ref{fig:vn_vs_Kn} are $v_3^{\rm hydro\!}/\varepsilon_3 = 0.2096$, $A_3 = 0.6973$, $B_3 = 1.7922$, and $C_3 = 13.945$.

As one can see, the simulations with the largest number of rescatterings are in the regime where $v_2$ and $v_3$ have almost reached their hydrodynamic values. 
The simulations with the middle $\langle\textrm{Kn}\rangle$ value are in the intermediate regime, while those with the largest $\langle\textrm{Kn}\rangle$ correspond to the few-collision case.
Note that for triangular flow $v_3$, whose development necessitates more rescatterings than $v_2$~\cite{Alver:2010dn,Kurkela:2020wwb} (although the rise with $\langle N_\textrm{resc.}\rangle$ seems to be steeper), our simulations with the intermediate value of the cross section are further away from the corresponding hydrodynamic value. 
In turn, in the simulations with the largest $\langle\textrm{Kn}\rangle$, $v_3$ is still at the subpercent level, corresponding rather to the far-from-equilibrium regime.

\section{Quantifying the flow fluctuations}
\label{sec:analysis}

Let us now describe the procedure we adopt to quantify the fluctuations of the anisotropic flow harmonics. 

In this theoretical study, we know the initial state of a given event, in particular the orientations $\Phi_n$ of the symmetry planes of the initial geometry, see Eq.~\eqref{eq:eccentricities}, relative to the direction of the impact parameter.
Similarly, we can meaningfully compute the flow coefficients and the associated symmetry plane azimuths event by event. 
More accurately they are computed by averaging over the $N_\textrm{iter.}$ iterations with the same initial energy-density profile, which as we argued above yields values of $v_n$ reliable at a few $10^{-4}$ level, see Eq.~\eqref{delta_vn}.

To exploit this wealth of information, we shall not study the traditional coefficients $v_n$ defined in Eq.~\eqref{eq:def_vn}, but rather separately the cosine and sine parts:
\begin{equation}
\label{def_v_n,c/s}
	v_{n,\mathrm{c}} \equiv \langle \cos(n\varphi) \rangle\quad,\quad
	v_{n,\mathrm{s}} \equiv \langle \sin(n\varphi) \rangle,
\end{equation}
where $\langle \cdots \rangle$ denotes an average over particles.
Obviously these coefficients obey
\begin{equation}
v_{n,\mathrm{c}} = v_n\cos(n\Psi_n)\quad,\quad
v_{n,\mathrm{s}} = v_n\sin(n\Psi_n),
\end{equation}
so that we can reconstruct $\Psi_n$ from the knowledge of both $v_{n,\mathrm{c}}$ and $v_{n,\mathrm{s}}$ in an event.

Consistent with our use of $v_{n,\mathrm{c}}$ and $v_{n,\mathrm{s}}$ --- which we shall collectively denote $v_{n,\mathrm{c/s}}$ ---, we characterize the initial state of an event by 
\begin{equation}
\label{def_eps_n,c/s}
	\varepsilon_{n,\mathrm{c}} \equiv -\frac{\langle r^n\cos(n\theta) \rangle}{\langle r^n \rangle}
	\quad,\quad
	\varepsilon_{n,\mathrm{s}} \equiv -\frac{\langle r^n\sin(n\theta) \rangle}{\langle r^n \rangle},
\end{equation}
where $\langle \cdots \rangle$ now denotes an average weighted with the energy density.
These ``eccentricities'' are clearly related to the real and imaginary parts of $\varepsilon_n$ defined in Eq.~\eqref{eq:eccentricities}.
In contrast to the convention for the latter, the coefficients $\varepsilon_{n,\mathrm{c/s}}$ are not necessarily positive.

\begin{figure}[tb!]
	\includegraphics*[width=\linewidth]{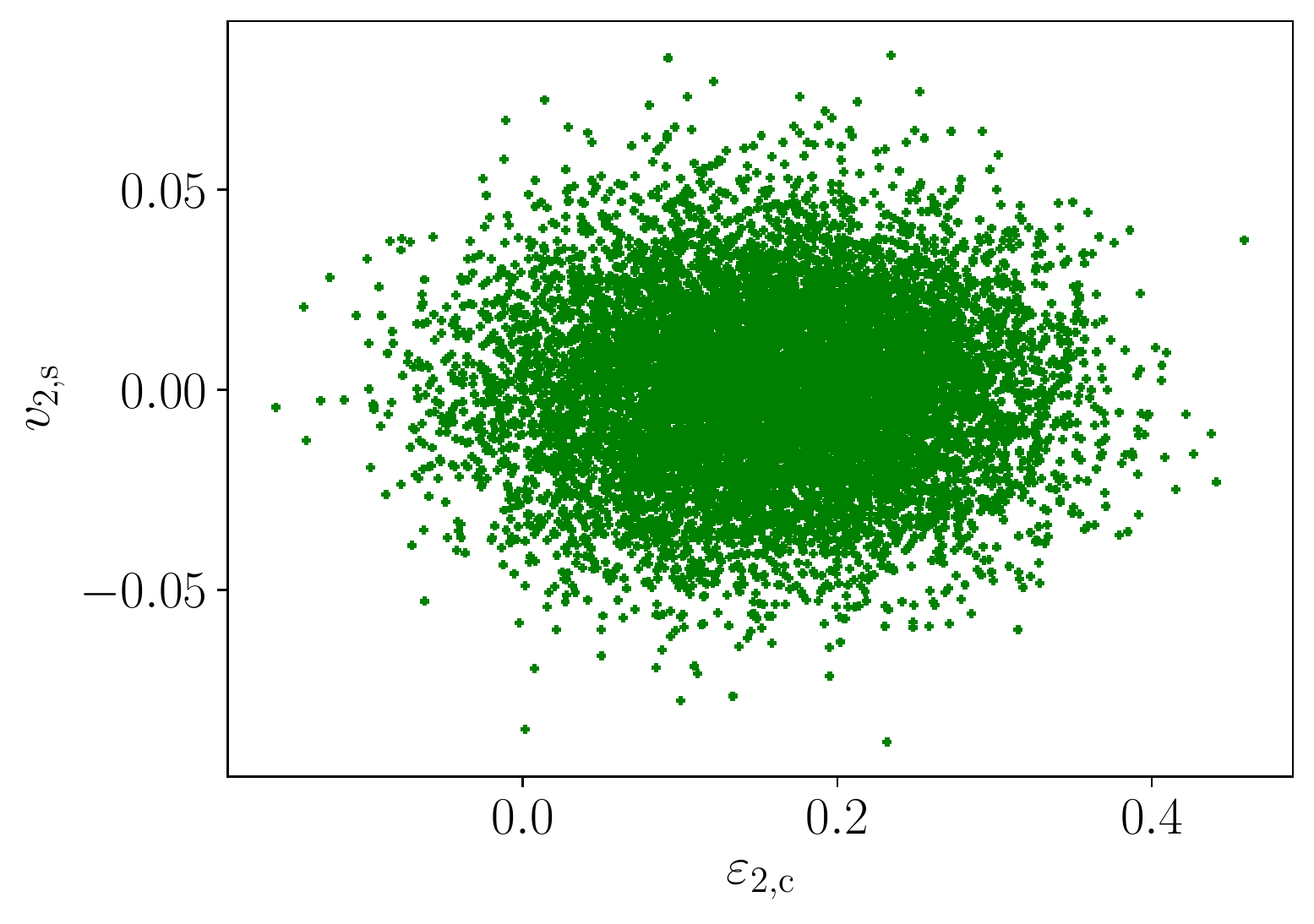}
	
	\caption{Scatter plot of the final $v_{2,\mathrm{s}}$ vs.\ the initial $\varepsilon_{2,\mathrm{c}}$ for events at $b=6$~fm with $\langle \textrm{Kn}\rangle^{-1}=13.74$.}
	\label{fig:v2s_vs_eps2c_b6}
\end{figure}

An advantage of investigating the behaviors of $v_{n,\mathrm{c}}$ and $v_{n,\mathrm{s}}$ separately is that their geometric causes $\varepsilon_{n,\mathrm{c}}$, $\varepsilon_{n,\mathrm{s}}$ are to a very large extent uncorrelated, as illustrated by the scatter plot of $v_{2,\rm s}$ (at the end of the evolution) as a function of $\varepsilon_{2,\rm c}$ shown in Fig.~\ref{fig:v2s_vs_eps2c_b6} for events at $b=6$~fm with the largest cross section we simulate.
Restricting ourselves to the lowest harmonics $n=2$ and 3, we thus already have four independently fluctuating initial-state eccentricities to our disposal, and as many anisotropic flow coefficients in the final state, whose fluctuations we can study as a function of the inverse mean Knudsen number $\langle \textrm{Kn}\rangle^{-1}$.
On the other hand, a drawback is that $v_{n,\mathrm{c}}$ and $v_{n,\mathrm{s}}$ are not measured in experiments. 
Yet as we already mentioned, our two-dimensional model is anyway not adequate for quantitative phenomenological studies.\footnote{Additionally, we work with a fixed impact parameter, which is also not feasible experimentally.} 

At a given impact-parameter value, we simulate $10^4$ initial energy density profiles. 
For each of those, we compute the four eccentricities $\varepsilon_{2,\mathrm{c}}$, $\varepsilon_{2,\mathrm{s}}$, $\varepsilon_{3,\mathrm{c}}$, $\varepsilon_{3,\mathrm{s}}$. 
Each energy density distribution is converted into $N_{\rm iter.} = 10$ particle distributions. 
These are then let to evolve with the transport algorithm, yielding at the end of the evolution flow coefficients $v_{2,\mathrm{c}}$, $v_{2,\mathrm{s}}$, $v_{3,\mathrm{c}}$, $v_{3,\mathrm{s}}$, which are averaged over the $N_{\rm iter.}$ runs with the same initial condition. 
Thus, we have to our disposal $10^4$ events with known initial eccentricities and final anisotropic flow coefficients, which we can correlate together. 

\begin{figure*}[ht!]
	\includegraphics*[width=.495\linewidth]{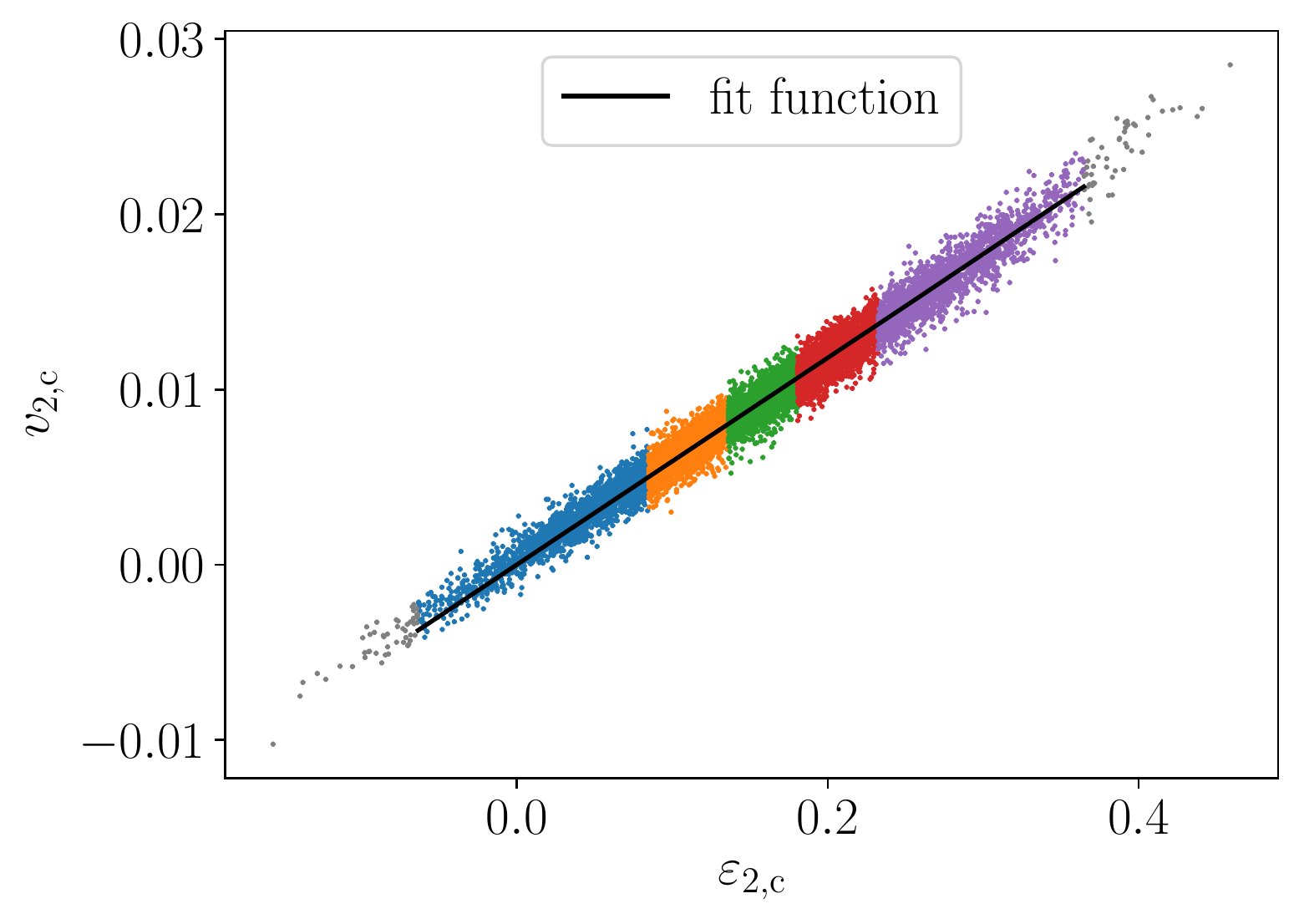}
	\includegraphics*[width=.495\linewidth]{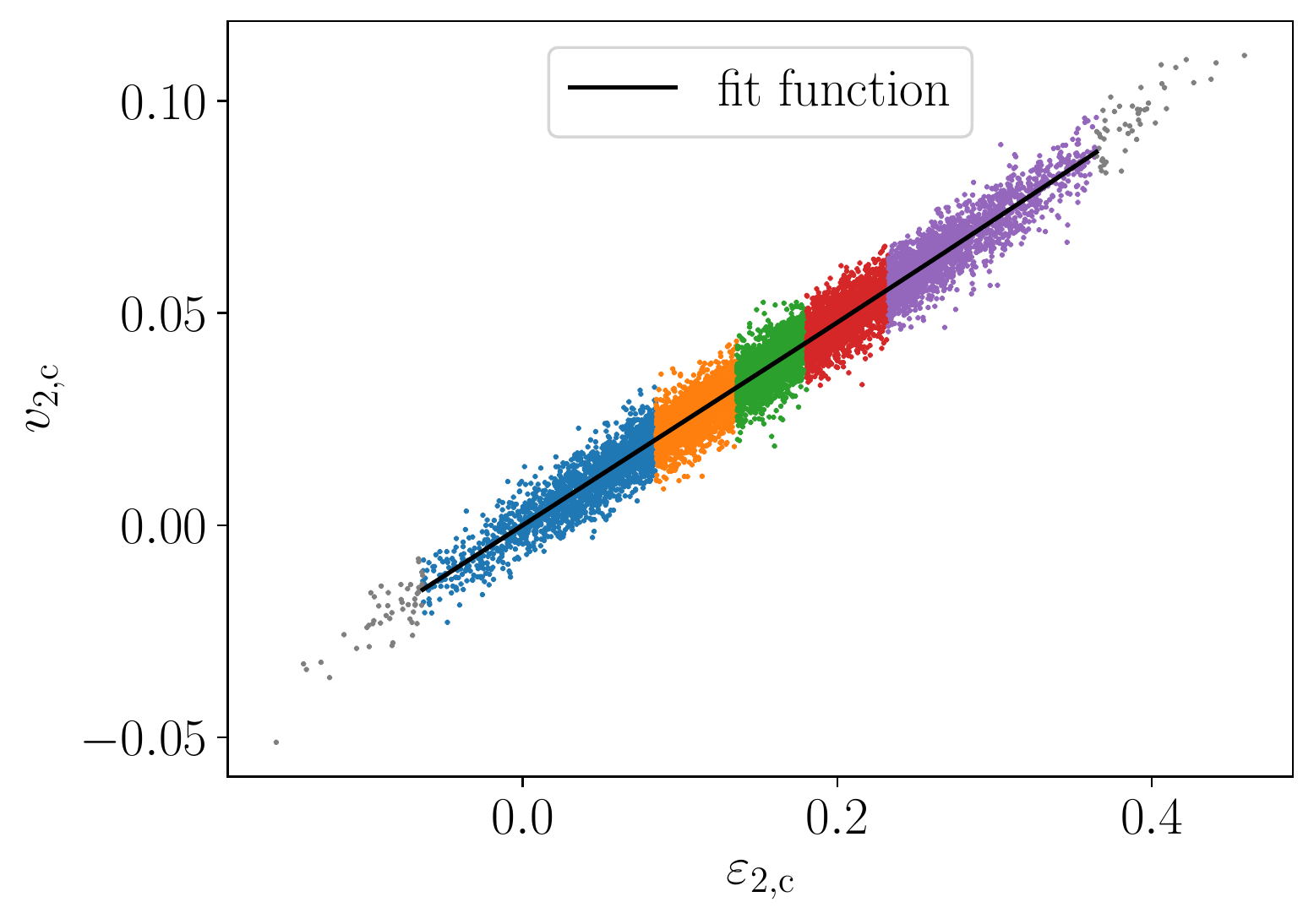}
	\caption{\label{fig:Scatter_b6}Scatter plot of the final $v_{2,\mathrm{c}}$ vs.\ $\varepsilon_{2,\mathrm{c}}$ for events at $b=6$~fm with $\langle \textrm{Kn}\rangle^{-1}=0.34$ (left) and $\langle \textrm{Kn}\rangle^{-1}=13.74$ (right).
	The full curves are fits with Eq.~\eqref{eq:v2_vs_eps2_nonlin}. 
	The grey points signal the events that are removed from our analysis, while the other colors correspond to the 5 bins in which the statistical properties are investigated independently (see main text).}
\end{figure*}
For each cosine or sine harmonic, we remove as a first step events with the smallest and largest eccentricities, 50 each, to get rid of outliers.\footnote{The main purpose here was to avoid having eccentricity bins that are too wide. We checked that our results are independent of how many events have been removed.} 
The remaining events are then divided according to their eccentricity value into five bins with 1950 events for the first and last bins and 2000 events for the central bins. 
This is illustrated by the different colors of the points in Fig.~\ref{fig:Scatter_b6}, in which we display $v_{2,\mathrm{c}}$ as a function of $\varepsilon_{2,\mathrm{c}}$ for collisions at $b=6$~fm for two values of the cross section.
We then perform a global fit using either a linear or a linear + cubic dependence, as in Eqs.~\eqref{eq:v2_vs_eps2_lin}--\eqref{eq:v2_vs_eps2_nonlin}.
The resulting fit function will be denoted by $\bar{v}_{n,\mathrm{c/s}}(\varepsilon_{n,\mathrm{c/s}})$. 
In practice, the fit with a cubic term was only relevant for the $v_{2,\mathrm{c}}$ and $v_{2,\mathrm{s}}$ at $b=6$ and 9~fm. 
The fit parameters for the four flow coefficients for the various sets of events we consider are listed in Appendix~\ref{app:fit-params}.
We also present in Appendix~\ref{app:complex_eps2,v2} the complex initial eccentricity $\varepsilon_{2\,}{\rm e}^{2{\rm i}\Phi_2}$ and final elliptic flow $v_{2\,}{\rm e}^{2{\rm i}\Psi_2}$ for the events of Fig.~\ref{fig:Scatter_b6}.

A comparison by eye between both panels of Fig.~\ref{fig:Scatter_b6} already hints at the larger width of the simulations with more rescatterings (right) compared to those with smaller inverse Knudsen number on the left. 
To quantify this observation, which also holds for the other flow harmonics, we calculate the moments of the distribution of $v_{n,\mathrm{c}/\mathrm{s}}$ about the fit function. 
More precisely, given the values of $\varepsilon_{n,\mathrm{c/s}}$ and $v_{n,\mathrm{c/s}}$ for an event, which we shall denote in the form $v_{n,\mathrm{c/s}}(\varepsilon_{n,\mathrm{c/s}})$, we compute powers of the difference $v_{n,\mathrm{c/s}}(\varepsilon_{n,\mathrm{c/s}})-\bar{v}_{n,\mathrm{c/s}}(\varepsilon_{n,\mathrm{c/s}})$, where the fit function is evaluated at the same value of the eccentricity.  
If we had infinite statistics, we would thus be computing the moments of the conditional probability distribution $p_{v|\varepsilon}(v_{n,\mathrm{c}/\mathrm{s}}|\varepsilon_{n,\mathrm{c}/\mathrm{s}})$ about the fit function.
Since we only have limited statistics, we can only approach the behavior of these conditional probabilities, which is why we use finite-width bins in $\varepsilon_{n,\mathrm{c}/\mathrm{s}}$. 
We do not gather all points in a single bin from the start to avoid averaging out opposite trends in different eccentricity intervals---which can indeed happen, as we shall discuss below.

Note that we cannot reasonably estimate $p_{v|\varepsilon}(v_{n,\mathrm{c/s}}|\varepsilon_{n,\mathrm{c/s}})$ by using the relation
\begin{equation}
	p_{v|\varepsilon}(v_{n,\mathrm{c/s}}|\varepsilon_{n,\mathrm{c/s}}) = 
	\frac{p_{v}(v_{n,\mathrm{c/s}})}{p_\varepsilon(\varepsilon_{n,\mathrm{c/s}})}
	\label{eq:cond_probab}
\end{equation}
in terms of the probability distributions for $v_{n,\mathrm{c/s}}$ and $\varepsilon_{n,\mathrm{c/s}}$ because of the division by very small numbers at the tails of the distributions that it would imply. 
Instead we tackle the left hand side of Eq.~\eqref{eq:cond_probab} directly.

Let us introduce the moments we shall report in Sec.~\ref{sec:results}, dropping momentarily the c/s indices for brevity.
The first moment about the fit function is 
\begin{equation}
\mu=\frac{1}{N}\sum_{i=1}^N \big[v_{n,i}(\varepsilon_{n,i})-\bar{v}_{n,i}(\varepsilon_{n,i})\big],
\label{eq:mean}
\end{equation}
where the index $i$ runs over all $N$ events in a given eccentricity bin.
We shall not report the value of $\mu$ in the following:  
it is always a small number of order $10^{-5}$, although it need not vanish in a bin. 
Yet its average over bins does vanish by construction of the fit function.

The second moment about the fit function, which almost coincides with the variance about the true average in a bin, is defined by
\begin{equation}
	\label{def_variance}
\sigma_v^2 = \frac{1}{N}\sum\limits_{i=1}^N \big[v_{n,i}(\varepsilon_{n,i})-\bar{v}_{n,i}(\varepsilon_{n,i})\big]^2.
\end{equation}
A trivial dependence on the inverse mean Knudsen number comes from the fact that the $v_n$ values themselves grow (in absolute value) with $\langle\textrm{Kn}\rangle^{-1}$. 
To account for this growth, we shall give values of $\sigma_v^2$ divided by the square $\mathcal{K}_{n,n}^2$ of the linear coefficient from the corresponding fit function. 

The third moment, or skewness, is given by
\begin{equation}
\gamma_1 = \frac{1}{N}\sum_{i=1}^N \frac{\big[v_{n,i}(\varepsilon_{n,i})-\bar{v}_{n,i}(\varepsilon_{n,i})\big]^3}{\sigma_v^3}.
\end{equation}
Since it involves an odd power it can be either positive or negative. 
Indeed both possibilities happen as we shall discuss in Sec.~\ref{ss:results_moments}, and it is precisely to account for this behavior that all events were not analyzed in a single bin. 

Eventually, we shall also present for a fourth-moment related quantity, namely the (excess) kurtosis
\begin{equation}
\label{def_kurtosis}
\gamma_2 = \frac{1}{N}\sum\limits_{i=1}^N \frac{\big[v_{n,i}(\varepsilon_{n,i})-\bar{v}_{n,i}(\varepsilon_{n,i})\big]^4}{\sigma_v^4} - 3.
\end{equation}
Note that while the early study~\cite{Vogel:2007yq} only investigated the variance of the conditional probability of elliptic flow $v_2$ at a fixed eccentricity $\varepsilon_2$, in this paper we go beyond the Gaussian approximation and study higher moments. 

We characterize the uncertainty on the value of a given moment ${\cal M}$ by a variance $\sigma_{\cal M}^2$.
The latter are obtained from a delete-d Jackknife algorithm with random sampling \cite{Jackknife}, where we randomly delete ten percent of the data in a bin and calculate the moments.
This is repeated for 3000 random samples and the sample variance is computed.

These moments~\eqref{def_variance}--\eqref{def_kurtosis} --- in the case of the skewness, its absolute value $|\gamma_1|$ --- and their respective variances are finally averaged over the 5 eccentricity bins.
For this, we use a weighted average to give less weight to the bins with larger uncertainties.
The weighted average $\langle {\cal M}\rangle$ of a given moment ${\cal M}$ and the associated variance $\sigma^2_{\cal M}$ are computed by
\begin{equation}
\langle {\cal M}\rangle = \frac{\sum_j {\cal M}_j/\sigma^2_{{\cal M},j}}{\sum_j 1/\sigma^2_{{\cal M},j}},
\end{equation}
\begin{equation}
\sigma_{\cal M}^2 = \frac{N_\mathrm{bins}}{\sum_j 1/\sigma^2_{{\cal M},j}},
\end{equation}
where the index $j$ runs over the $N_\mathrm{bins}$ bins, ${\cal M}_j$ is the value of a moment in bin $j$, and $\sigma_{{\cal M},j}^2$ the variance $\sigma_{\cal M}^2$ in that bin.

\section{Results}
\label{sec:results}

Let us now present our results for the behavior of the flow harmonics as a function of the number of rescatterings in the system.  
In this section we only give a representative sample from our results at the three different impact parameters, $b=0$, 6, and 9~fm.
Further results are given in the appendices.

\subsection{Initial anisotropic flow}
\label{ss:results_initial}

\begin{figure*}[ht!]
	\includegraphics*[width=.495\linewidth]{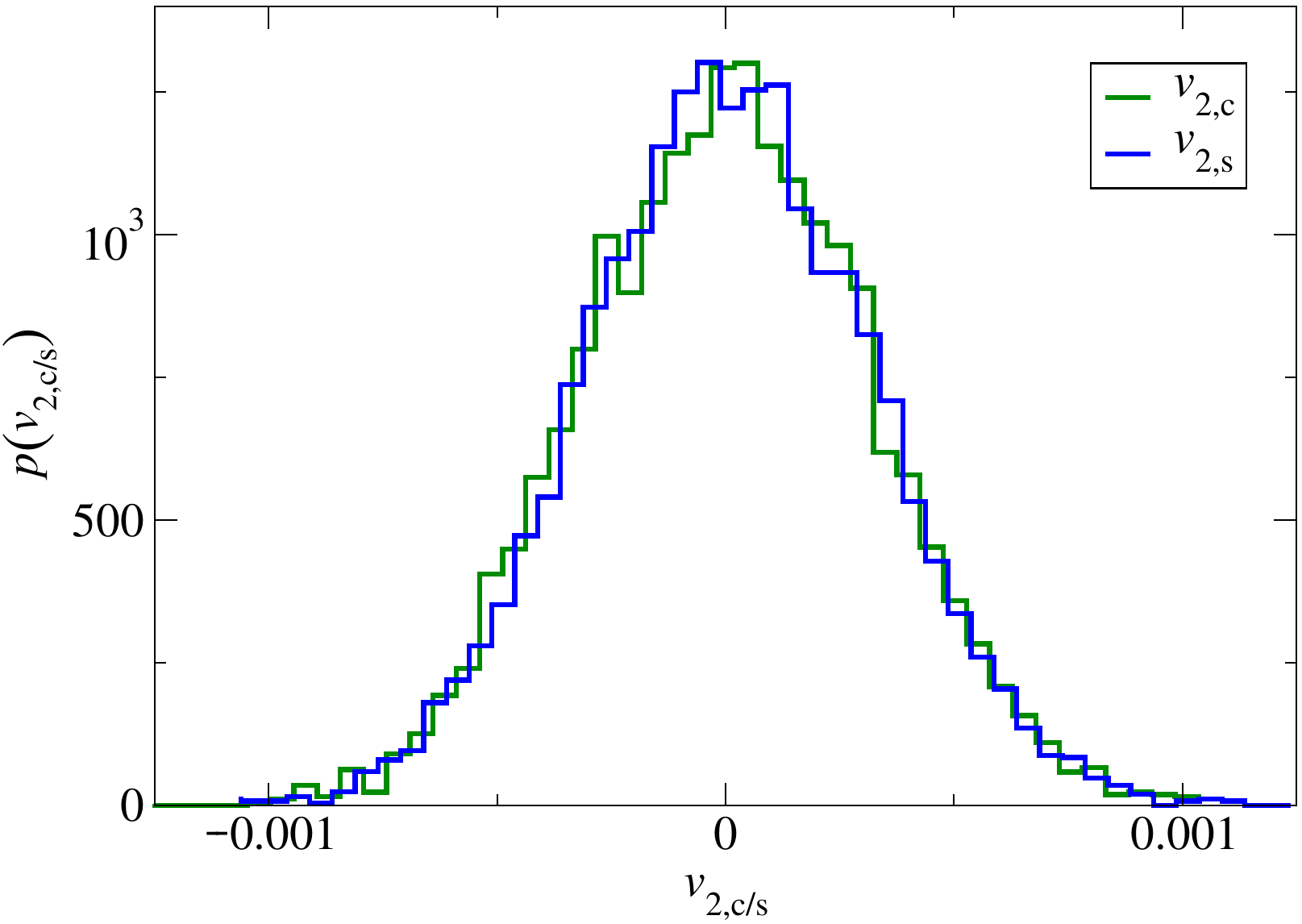}
	\includegraphics*[width=.495\linewidth]{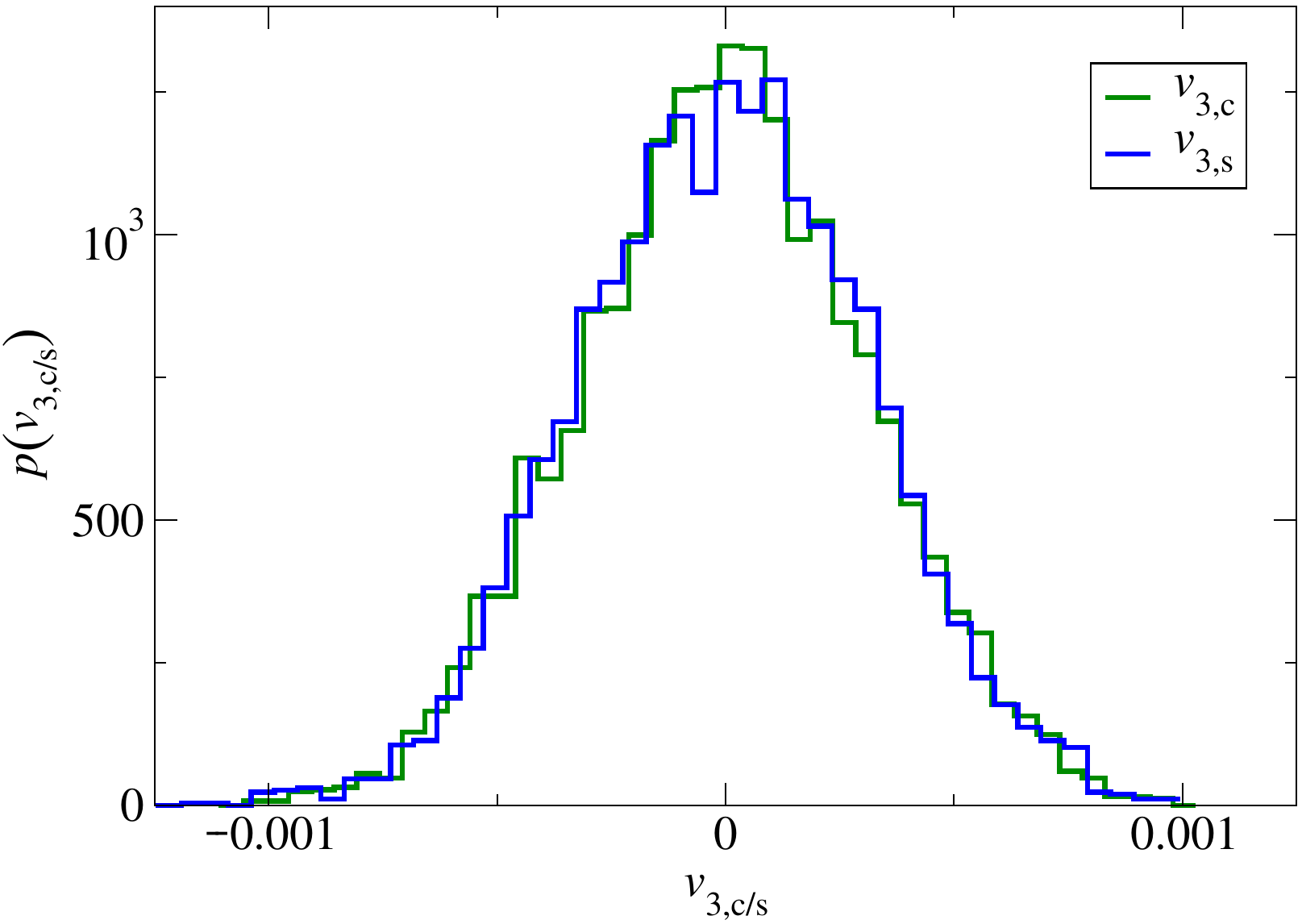}
	\caption{\label{fig:initial-vn_hist_b0}Probability distribution of $v_{2,\mathrm{c}/\mathrm{s}}$ (top) and  $v_{3,\mathrm{c}/\mathrm{s}}$ (bottom) in the initial state of the transport algorithm computed for 5000 events at $b=0$ fm.}
\end{figure*}

Since the number of particles $N_{\rm p}$ in each event is finite, the initial momentum distribution before any evolution may already show some sizable anisotropic flow: as a rough estimate, one can anticipate that the initial values of $v_{n,\rm c}$ or $v_{n,\rm s}$ should fluctuate about 0 with a dispersion of order $1/\sqrt{2N_{\rm p}} = 10^{-3}$~\cite{Voloshin:1994mz}.\footnote{This does not hold for $n=1$ since we enforced a vanishing total momentum in the initial state.}
More precisely, these numerical initial-state fluctuations of $v_{n,\rm c/s}$ should even be Gaussian distributed, and the event-plane angle $\Psi_n$ should be uncorrelated to the participant-plane orientation $\Phi_n$.

To test these expectations, we chose 5000 among the $10^4$ Monte Carlo initial conditions that are simulated for each of the three impact parameter values we consider.
We computed the corresponding initial flow coefficients $v_{2,\rm c}$, $v_{2,\rm s}$, $v_{3,\rm c}$, $v_{3,\rm s}$ in each event, which gives us the approximate probability distribution of these harmonics --- irrespective of the corresponding spatial eccentricities. 
As an example, histograms for the probability distributions obtained at $b=0$ are shown in Fig.~\ref{fig:initial-vn_hist_b0}.

To characterize these probability distributions, we computed their first moments performing an analysis similar to that described in Sec.~\ref{sec:analysis}, with  a few differences.
First, we used a single bin for all events and thus analyze of all them at once irrespective of their eccentricities, retaining all events. 
Then we directly assumed that the average values of $v_{n,\rm c/s}$ are zero (the actual mean values are of order $10^{-5}$) when computing the moments.
Finally, in the Jackknife algorithm for the uncertainties on the moments we deleted 500 events from each set, and repeated the calculations over 6000 samples.
The resulting variance $\sigma_v^2$, skewness (in absolute value) $\gamma_1$, and excess kurtosis $\gamma_2$ are given in Table~\ref{tab:initial-vn_b0}, in which we actually scaled $\sigma_v^2$ by its naive estimate $1/2N_{\rm p}$ for fluctuations of purely numerical origin.
We indeed find $\sigma_v\simeq 1/\sqrt{2N_{\rm p}}$, while the skewness and excess kurtosis are compatible with zero, i.e.\ the values for Gaussian distributions. 

\begin{table}[h]
	\caption{\label{tab:initial-vn_b0}Moments of the probability distributions displayed in Fig.~\ref{fig:initial-vn_hist_b0} for the anisotropic flow harmonics in the initial state computed for 5000 events at $b=0$~fm. 
		The variance $\sigma_v^2$ is divided by the estimate $1/2N_{\rm p}$.}
	\begin{ruledtabular}
		\begin{tabular}{cccc}
			& $\sigma_v^2/(2N_{\rm p})^{-1}$ & $|\gamma_1|$ & $\gamma_2$ \\ \hline
			$v_{2,c}$ & $1.014 \pm 0.020$ & $0.041 \pm 0.055$ & $-0.030 \pm 0.069$ \\
			$v_{2,s}$ & $0.990 \pm 0.020$ & $0.079 \pm 0.057$ & $0.063 \pm 0.078$ \\
			$v_{3,c}$ & $0.996 \pm 0.020$ & $0.034 \pm 0.057$ & $0.013 \pm 0.075$ \\
			$v_{3,s}$ & $0.996 \pm 0.020$ & $0.089 \pm 0.055$ & $-0.014 \pm 0.073$ \\ 
		\end{tabular}
	\end{ruledtabular}
\end{table}

Similar-looking histograms, leading to comparable moments of the probability distributions, are found for $b=6$ or 9~fm (see Appendix~\ref{app:initial-vn_b9}), which shows that these initial flow harmonics are driven by numerical fluctuations only.

\subsection{Moments of the final state distribution}
\label{ss:results_moments}

We now turn to the flow harmonics in the final state, as a function of  $\langle \textrm{Kn}\rangle^{-1}$.
Before we show any result, we emphasize that the trends they exhibit are consistent across different impact parameters.
Accordingly, here and in Sec.~\ref{ss:result_angles} we only show results at $b=6$~fm, postponing those at 0 or 9~fm to Appendices~\ref{app:moments_b0-9} and \ref{appendix:rp_dist}. 

\begin{table*}[tb!]
	\caption{\label{tab:final-v2_b6}Moments of the conditional probability distributions $p_{v|\varepsilon}(v_{2,\rm c/s}|\varepsilon_{2,\rm c/s})$ of the final-state flow harmonics computed for $10^4$ events at $b=6$~fm. 
		The variance $\sigma_v^2$ is divided by the square ${\cal K}_{2,2}^2$ of the linear coefficient in relation~\eqref{eq:v2_vs_eps2_nonlin}.}
	\begin{ruledtabular}
		\begin{tabular}{ccccccc}
			$\langle{\rm Kn}\rangle^{-1}$ & \multicolumn{3}{c}{$v_{2,\rm c}$}  & \multicolumn{3}{c}{$v_{2,\rm s}$} \\
			& $10^4\,\sigma_v^2/{\cal K}_{2,2}^2$ & $|\gamma_1|$ & $\gamma_2$ & 
			   $10^4\,\sigma_v^2/{\cal K}_{2,2}^2$ & $|\gamma_1|$ & $\gamma_2$\\ \hline
			0.34 & $1.754 \pm 0.062$ & $0.513 \pm 0.091$ & $0.317 \pm 0.141$ & 
              $1.593 \pm 0.055$ & $0.249 \pm 0.102$ & $0.302 \pm 0.165$ \\
			3.43 & $1.313 \pm 0.046$ & $0.281 \pm 0.105$ & $0.481 \pm 0.188$ & 
			  $1.327 \pm 0.046$ & $0.257 \pm 0.100$ & $0.287 \pm 0.158$ \\
			13.73 & $3.185 \pm 0.104$ & $0.138 \pm 0.093$ & $0.080 \pm 0.135$ & 
              $3.150 \pm 0.104$ & $0.155 \pm 0.094$ & $0.088 \pm 0.132$ \\
		\end{tabular}
	\end{ruledtabular}
\end{table*}

\begin{table*}[tb!]
	\caption{\label{tab:final-v3_b6}Moments of the conditional probability distributions $p_{v|\varepsilon}(v_{3,\rm c/s}|\varepsilon_{3,\rm c/s})$ of the final-state flow harmonics computed for $10^4$ events at $b=6$~fm. 
		The variance $\sigma_v^2$ is divided by the square ${\cal K}_{3,3}^2$ of the linear coefficient in relation~\eqref{eq:v3_vs_eps3_lin}.}
	\begin{ruledtabular}
		\begin{tabular}{ccccccc}
			$\langle{\rm Kn}\rangle^{-1}$ & \multicolumn{3}{c}{$v_{3,\rm c}$}  & \multicolumn{3}{c}{$v_{3,\rm s}$} \\
			& $10^3\,\sigma_v^2/{\cal K}_{3,3}^2$ & $|\gamma_1|$ & $\gamma_2$ & 
			$10^3\,\sigma_v^2/{\cal K}_{3,3}^2$ & $|\gamma_1|$ & $\gamma_2$\\ \hline
			0.34 & $1.984 \pm 0.063$ & $0.282 \pm 0.086$ & $-0.026 \pm 0.109$ & 
			$2.480 \pm 0.079$ & $1.077 \pm 0.060$ & $0.026 \pm 0.113$ \\
			3.43 & $0.294 \pm 0.010$ & $0.183 \pm 0.103$ & $0.305 \pm 0.170$ & 
			$0.355 \pm 0.012$ & $0.576 \pm 0.098$ & $0.428 \pm 0.173$ \\
			13.73 & $0.697 \pm 0.022$ & $0.105 \pm 0.088$ & $-0.020 \pm 0.106$ & 
			$0.775 \pm 0.025$ & $0.258 \pm 0.085$ & $-0.017 \pm 0.097$ \\
		\end{tabular}
	\end{ruledtabular}
\end{table*}

In Table~\ref{tab:final-v2_b6} resp. Table~\ref{tab:final-v3_b6} we give the moments~\eqref{def_variance}--\eqref{def_kurtosis} for $v_{2,\rm c/s}$ resp.\ $v_{3,\rm c/s}$ for the three values of the cross section in our simulations.
Regarding the variance, here divided by ${\cal K}_{n,n}^2$ which monotonously increases with $\langle \textrm{Kn}\rangle^{-1}$, we observe that it is consistently highest for the simulations with the largest mean number of rescatterings. 
On the other hand, the values of $\sigma_v^2$ in the collisions with the smallest $\langle \textrm{Kn}\rangle^{-1}$ are close to that in the initial state and due to numerical noise (Sec.~\ref{ss:results_initial}). 
The overall trend seems therefore to be that the dispersion of the flow harmonics about their ``average'' value as represented by the fit function $\bar{v}_{n,\rm c/s}$ is increasing with $\langle \textrm{Kn}\rangle^{-1}$, and that this increase is stronger than that of $\bar{v}_{n,\rm c/s}$ itself. 
This corresponds to the visual impression of a larger width of the scatter plot in the right panel of Fig.~\ref{fig:Scatter_b6} mentioned above. 

Let us now discuss the skewness. 
In the tables, we only report absolute values, but $\gamma_1$ can be either positive or negative. 
Before averaging over the eccentricity bins, we actually observed that the sign of $\gamma_1$ does depend on the bin, with a clear trend. 
In bins with a positive $\varepsilon_{n,\rm c/s}$, resulting mostly in a positive final-state $v_{n,\rm c/s}$, the skewness $\gamma_1$ is negative. 
This signals a distribution with a longer tail towards the left of its positive average, i.e.\ a distribution skewed towards 0. 
This behavior was observed for the fluctuations of elliptic flow $v_2$ in hydrodynamics calculations~\cite{Giacalone:2016eyu}, but as we discuss in next section, the underlying reason probably differs. 
In turn, in bins with a negative eccentricity we find a positive skewness, which again means a distribution skewed towards 0.
To remedy this change of sign of $\gamma_1$, we average the absolute values $|\gamma_1|$. 
In most cases, the average $|\gamma_1|$ decreases with increasing $\langle \textrm{Kn}\rangle^{-1}$.
For the largest cross section we have studied, $|\gamma_1|$ seems to be approaching 0, i.e.\ the value for a Gaussian distribution.
On the other hand, in the simulations with the smallest $\langle \textrm{Kn}\rangle^{-1}$ it is far from 0 --- and in particular much larger than in the initial state. 
This hints at the fact that a (very) small number of rescatterings\footnote{Unfortunately, smaller than in the simulations we made.} would lead to a $v_{n,\rm c/s}$ distribution with a non-zero mean and a strong skewness towards zero, and that $|\gamma_1|$ then decreases with increasing $\langle \textrm{Kn}\rangle^{-1}$.

Regarding the excess kurtosis $\gamma_2$, the rather large error bars make it difficult to draw definite conclusions. 
If one nevertheless looks for overall trends, one can identify two. 
First, $\gamma_2$ seems to favor positive values, and often values that are not compatible with 0. 
If we try to be more specific, one could say that $\gamma_2$ is always positive for $v_2$, while it is at times negative (yet compatible with 0 with error bars) for $v_3$, which happens to coincide with the findings in Ref.~\cite{Bhalerao:2018anl} in a hydrodynamic approach.
A second possible trend is a non-monotonic behavior with $\langle \textrm{Kn}\rangle^{-1}$, with a positive maximum --- signaling a distribution that is more peaked than a Gaussian --- at the intermediate value that we simulated, while for the smallest and largest $\langle \textrm{Kn}\rangle^{-1}$ the behavior is more Gaussian. 

In general we can say that for many collisions the absolute value of the skewness and the kurtosis approach the values for an almost Gaussian shape, as in the initial state.
Regarding the variance of the final-state flow harmonics, it is larger than in the initial state, 
Eventually, the general effect of considering a finite number of rescatterings seems be a larger absolute skewness than in the hydrodynamic case, as well as a larger kurtosis.

\subsection{Event-plane angle distributions}
\label{ss:result_angles}

\begin{figure*}[tb!]
	\centering
	\includegraphics*[width=.495\linewidth]{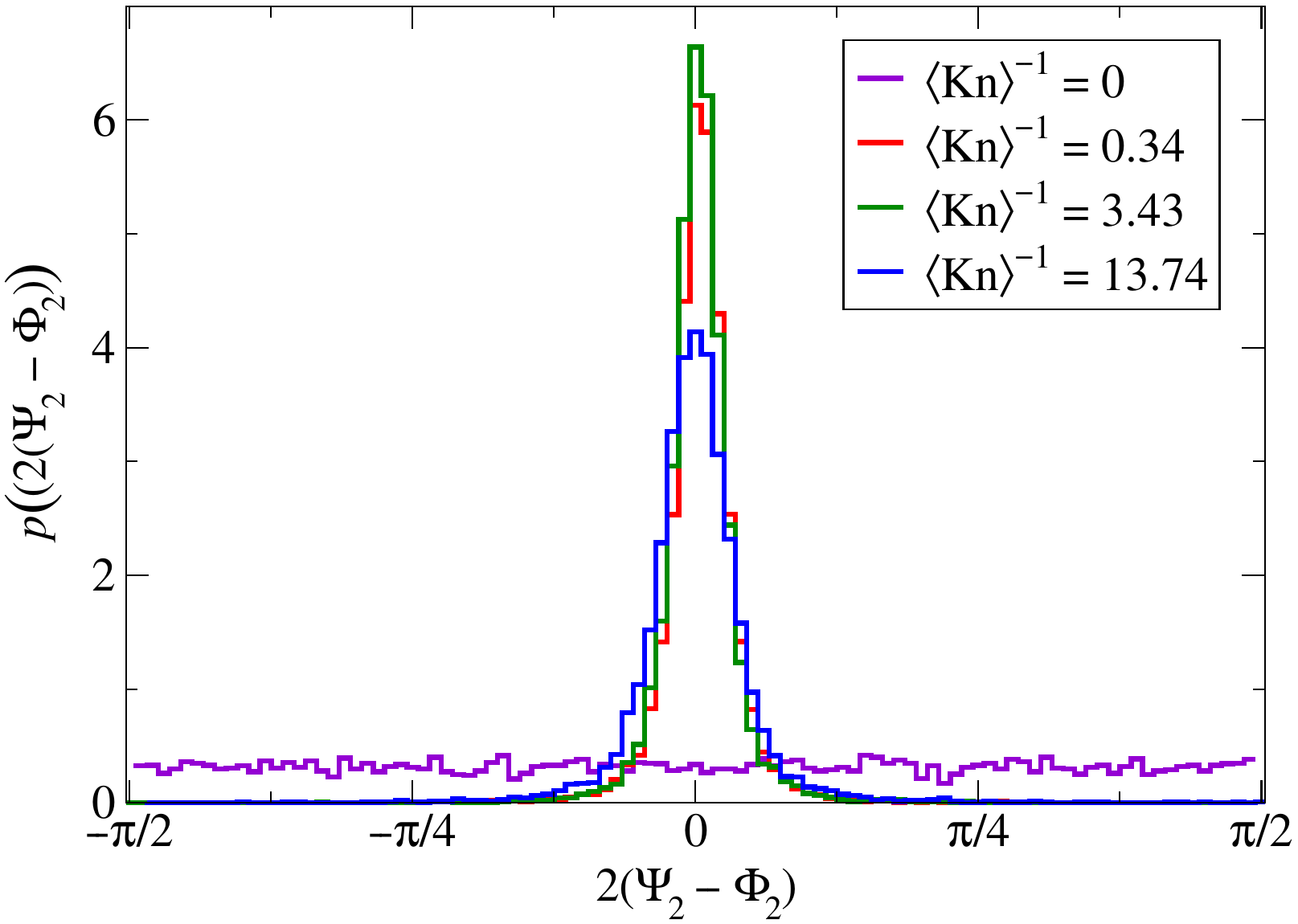}
	\includegraphics*[width=.495\linewidth]{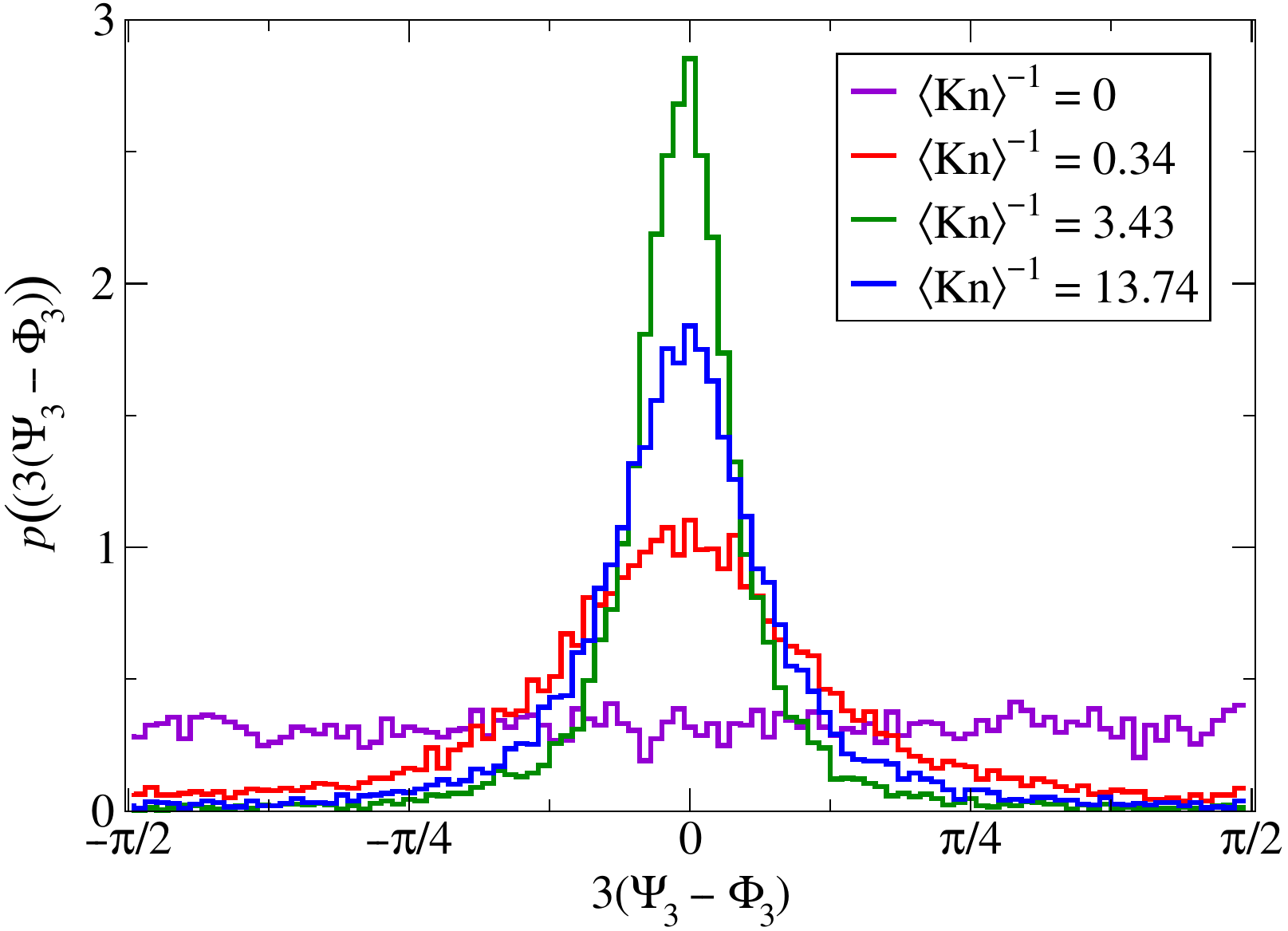}
	\caption{Probability distributions of $p(2(\Psi_2-\Phi_2))$ (left) and  $p(3(\Psi_3-\Phi_3))$ (right) for events at $b=6$ fm for different $\langle \textrm{Kn}\rangle^{-1}$.
	The value $\langle \textrm{Kn}\rangle^{-1}=0$ corresponds to the initial state.}
	\label{fig:b6_angles}
\end{figure*}

From the cosine and sine parts of a flow harmonic, we can calculate the corresponding flow angle $\Psi_n$. 
Similarly from $\varepsilon_{n,\mathrm{c}}$ and $\varepsilon_{n,\mathrm{s}}$ we extract the orientation $\Phi_n$ of the $n$-th symmetry plane in the initial state. 
Therefore, we can study probability distributions $p(n(\Psi_n-\Phi_n))$ for $n = 2$ and 3.
These are displayed in Fig.~\ref{fig:b6_angles} for our simulations at $b=6$~fm, as a function of $\langle \textrm{Kn}\rangle^{-1}$. 
The results for the simulations at 0 or 9~fm are in Figs.~\ref{fig:b0_angles}, \ref{fig:b9_angles} in Appendix~\ref{appendix:rp_dist}.
In the plots, we also show the distributions about the respective participant plane of the flow angles $\Psi_2$ and $\Psi_3$ in the initial state of our simulations, with the label $\langle \textrm{Kn}\rangle^{-1}=0$. 
These initial distributions are flat, up to fluctuations, which shows that there is no correlation between the initial flow, due to numerical noise, in our simulations and the geometry of the events. 

Regarding the distributions of $\Psi_n-\Phi_n$ in the case when the flow angle $\Psi_n$ is that at the end of the evolution, let us discuss first the results for $n=3$. 
At the smallest value of $\langle \textrm{Kn}\rangle^{-1} \approx 0.3$--0.4, a still rather broad distribution develops with a maximum at $\Psi_3-\Phi_3 = 0$. 
This clearly signals the onset of the transfer from the initial anisotropy in position space into a signal in momentum space. 
At the 10 times larger $\langle \textrm{Kn}\rangle^{-1}$, the distribution is now very peaked around $\Psi_3-\Phi_3 = 0$: the flow angle is almost aligned along the symmetry plane of the initial geometry. 
Eventually, the distribution at the largest value of $\langle \textrm{Kn}\rangle^{-1}$ is less peaked, with thicker tails at larger values of $|\Psi_3-\Phi_3|$. 
We shall come back in next section to this behavior, which was not anticipated but is consistently observed at all three values of the impact parameter, and also in the second harmonic as we discuss now. 

Coming to the case $n=2$, let us recall our comment at the end of Sec.~\ref{sec:transport} regarding the slower onset of $v_3$ compared to $v_2$ as a function of $\langle \textrm{Kn}\rangle^{-1}$~\cite{Alver:2010dn,Kurkela:2020wwb}.
This behavior, which was known for the values of the flow coefficients, is also visible at the level of the flow angles.
Indeed, at the smallest value of $\langle \textrm{Kn}\rangle^{-1}$, we find that the distribution of $\Psi_2-\Phi_2$ is already very peaked around 0, much more so than $p(3(\Psi_3-\Phi_3))$. 
This is true not only at $b=6$ or 9~fm, in which the initial-state geometry yields a clear marked $\Phi_2$, but even (although to a lesser extent) at $b=0$~fm. 
As the number of rescatterings in the system or equivalently $\langle \textrm{Kn}\rangle^{-1}$ increases, the peak persists. 
Yet at the highest $\langle \textrm{Kn}\rangle^{-1}$, we again observe some broadening of the distribution, as in the case $n=3$. 

For the values of $\langle \textrm{Kn}\rangle^{-1}$ for which the peak is clearly marked, its width is of the same magnitude as reported in hydrodynamic~\cite{Qiu:2011iv} or hybrid~\cite{Petersen:2010cw} simulations.

Finally, let us mention that none of the distribution $p(n(\Psi_n-\Phi_n))$ can be reasonably fitted by a (truncated) Gaussian nor by a beta distribution.

\section{Discussion}
\label{sec:summary}

We have investigated the fluctuations of anisotropic flow in a two-dimensional system of massless particles undergoing elastic binary scatterings, as a function of the mean number or collisions per particle --- which we vary by choosing various values of the scattering cross section. 
More specifically, we study how each flow coefficient fluctuates about the ``preferred'' value for the corresponding eccentricity, for instance, how the cosine coefficient $v_{2,\rm c}$ fluctuates event-to-event about the value $\bar{v}_{2,\rm c}(\varepsilon_{2,\mathrm{c}})$ defined by the best fit to the scatter plot of $v_{2,\rm c}$ vs.\ $\varepsilon_{2,\mathrm{c}}$ with Eq.~\eqref{eq:v2_vs_eps2_nonlin} or its linear version. 
These fluctuations are characterized by successive moments (variance, skewness, excess kurtosis). 
In addition, we studied the distribution of the event plane $\Psi_2$ or $\Psi_3$ about the corresponding participant plane $\Phi_2$ resp.\ $\Phi_3$ in the initial state.

Let us gather the salient findings that appear to be robust across the different impact-parameter values we have studied. 
\begin{itemize}
\item The variance $\sigma_v^2$ of the fluctuations of $v_{n,\rm c/s}$ about the mean value $\bar{v}_{n,\rm c/s}(\varepsilon_{n,\mathrm{c/s}})$, scaled by the overall size of the flow harmonic, first decreases with $\langle\mathrm{Kn}\rangle^{-1}$ and increases again.

\item The distributions of $v_{n,\rm c/s}$ are always skewed towards 0. 
The skewness first jumps to a sizable value as flow sets on --- to fix ideas, $|\gamma_1|$ is of order 0.5 ---, then decreases again as the number of rescatterings increases. 

\item The excess kurtosis $\gamma_2$ is predominantly positive, i.e.\ the distribution is more peaked than a Gaussian with the same variance. 
It tends towards 0 with increasing $\langle\mathrm{Kn}\rangle^{-1}$.

\item The distribution of event plane angles about the orientation of the corresponding participant plane are peaked. 
The peak broadens for the largest value of $\langle\mathrm{Kn}\rangle^{-1}$.
\end{itemize}

It seems to us that a consistent picture emerges from these features. 
First, let us emphasize that the fluctuations we discussed above are not the usual fluctuations arising from those of the eccentricity in a centrality bin. 
They are fluctuations at fixed eccentricity --- which is why we referred to the conditional probability distribution $p_{v|\varepsilon}(v_{n,\mathrm{c}/\mathrm{s}}|\varepsilon_{n,\mathrm{c}/\mathrm{s}})$ ---, whose moments are averaged over the range of eccentricities found in collisions at a given impact parameter, to increase their significance. 
That is, we are looking at fluctuations in the response function of anisotropic flow to the initial eccentricity~\cite{Yan:2014nsa}.
The physical origin of these fluctuations is essentially the finite number of scatterings that build up the anisotropic flow from the initial eccentricity of the geometry.
In turn, the value about which a given flow coefficient fluctuates is that which would be obtained when solving the (fully deterministic) kinetic Boltzmann equation at the same Knudsen number or equivalently opacity~\cite{Kurkela:2020wwb,Kurkela:2018qeb}.%
\footnote{Here we implicitly assume that the Boltzmann equation in terms of a smooth single-particle phase-space distribution is the actual evolution equation describing the system dynamics, which should hold in the dilute regime.}
Similarly, the fluctuations we have studied are absent from hydrodynamic simulations, at least of those implementing non-fluctuating hydrodynamics.

That being told, how can we interpret the trends in the moments of about the mean value?
The first feature is that the fluctuations from the finite number of rescatterings are non-Gaussian, meaning that the central limit theorem does not apply, precisely because there are not enough scatterings. 
The generic skewness towards zero seems to be intuitively clear, and comes from the fact that the initial value of each flow coefficient is zero and that $v_{n,\rm c/s}$ is bounded in absolute value.
On the other hand, the origin of the peaking of the distribution encoded in the kurtosis is not clear to us. 
Yet that this positive excess kurtosis tends to disappear in the collisions with the largest cross section seems to us to be correlated to the second main feature we observed, namely the broadening of the $\Psi_n-\Phi_n$ distributions in the same regime, which we also find consistently but do not really understand.
A possibility --- that we have not further tested --- is that for the largest $\langle\mathrm{Kn}\rangle^{-1}$ nonlinear ``contamination'' from other harmonics start to be sizable, as e.g.\ a contribution $\varepsilon_2\varepsilon_4$ to (the fluctuations of) $v_2$, both in value and orientation.

The present study can in natural way be extended in several directions. 
First, one can think of looking at simulations with either less or more rescatterings per particle. 
Less rescatterings, i.e.\ a larger Knudsen number, would allow one to study the onset of the skewness, or of the alignment of the event plane angle $\Phi_2$ along the participant plane orientation $\Psi_2$. 
The issue is that less rescatterings mean a smaller flow signal, so one needs to increase the statistics to recover meaningful results. 
In the other limit, simulations with a smaller Knudsen number would allow to probe the ``hydrodynamic limit'' more carefully: 
do the fluctuations of $v_{n,\rm c/s}$ about their mean value due to the finite number of scatterings become smaller --- which would be our expectation?
How do the distributions of $\Psi_n-\Phi_n$ evolve? 
The issue with those simulations is that one must ensure that the system remains in the dilute regime, so that cranking up the cross section is not enough, but should be accompanied by a decrease in the number of test particles. 
In turn, this has to be compensated by an increase in the number of simulated events, to recover small error bars. 

Another direction is to look at higher flow harmonics like $v_4$ or $v_5$, or at the correlations between the fluctuations of different harmonics~\cite{ALICE:2016kpq}. 
Here again the signals will be smaller, thus statistics-costly. 
In addition, these higher harmonics are significantly affected by lower-order eccentricities~\cite{Borghini:2005kd,Gardim:2011xv,Niemi:2012aj,Teaney:2012ke,Borghini:2018xum}, which also begs for more statistics to disentangle the various influences. 
Eventually, instead of looking only at the average flow coefficients, one can investigate how transverse momentum dependent $v_n(p_T)$ are affected by the number of rescatterings.

Going beyond our fixed impact-parameter study, one needs to turn to fixed-centrality studies, in which the fluctuations due to the finite number of rescatterings combine with eccentricity fluctuations. 
This is probably the simplest extension of our work, and a first step towards investigations that can meaningfully be used for phenomenological comparisons with experimental data.  
Anticipating on these more realistic studies, one can reasonably imagine that the convolution of the dispersion of the eccentricities within a centrality bin on the one hand, and the fluctuation of the anisotropic-flow response due to the final number of rescatterings will lead to a distortion of the non-Gaussianities of the flow-fluctuation pattern compared to the underlying eccentricity fluctuations. 
In particular, since we consistently found that the distribution of a flow harmonic at fixed eccentricity is skewed towards zero, one can expect that the distributions of $v_2$ or $v_3$ in a centrality bin should be more skewed towards zero than the distributions of $\varepsilon_2$ or $\varepsilon_3$. 
Since the average number of rescatterings decreases as the collisions become more peripheral, this ``extra negative skewness'' of the probability distribution of $v_n$ should be more pronounced in the peripheral bins than in the central ones, although a quantitative statement necessitates more realistic computations.

\begin{acknowledgments}
The authors acknowledge support by the Deutsche Forschungsgemeinschaft (DFG, German Research Foundation) through the CRC-TR 211 'Strong-interaction matter under extreme conditions' - project number 315477589 - TRR 211.
Computational resources have been provided by the Bielefeld GPU Cluster and the Center for Scientific Computing (CSC) at the Goethe-University of Frankfurt.
We would also like to thank Hannah Elfner for fruitful discussions and providing computational resources from her projects.
\end{acknowledgments}

\appendix

\begin{widetext}

\section{Initial anisotropic flow}
\label{app:initial-vn_b9}

In this Appendix, we present for the sake of completeness the probability distribution of each $v_{n,\rm c/s}$ in the initial state for our sets of events at $b=6$ and 9~fm.
The histograms for the distributions are displayed in Figs.~\ref{fig:initial-vn_hist_b6}, \ref{fig:initial-vn_hist_b9}, and their first moments are listed in Tables~\ref{tab:initial-vn_b6}, \ref{tab:initial-vn_b9}.
Note that the distributions are extremely similar to those at $b=0$~fm presented in Sec.~\ref{ss:results_initial}.

\begin{figure*}[ht]
	\includegraphics*[width=0.495\linewidth]{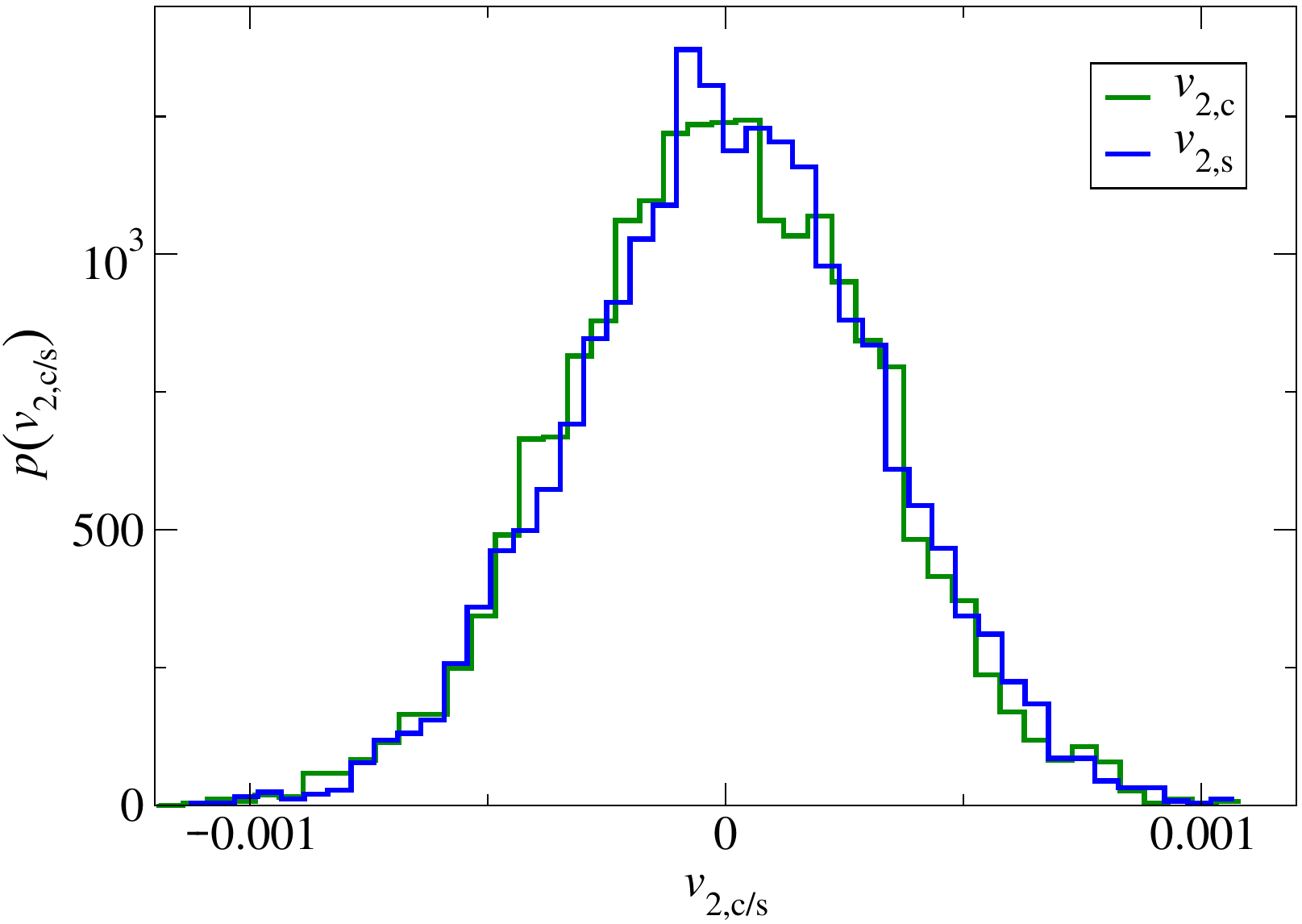}
	\includegraphics*[width=0.495\linewidth]{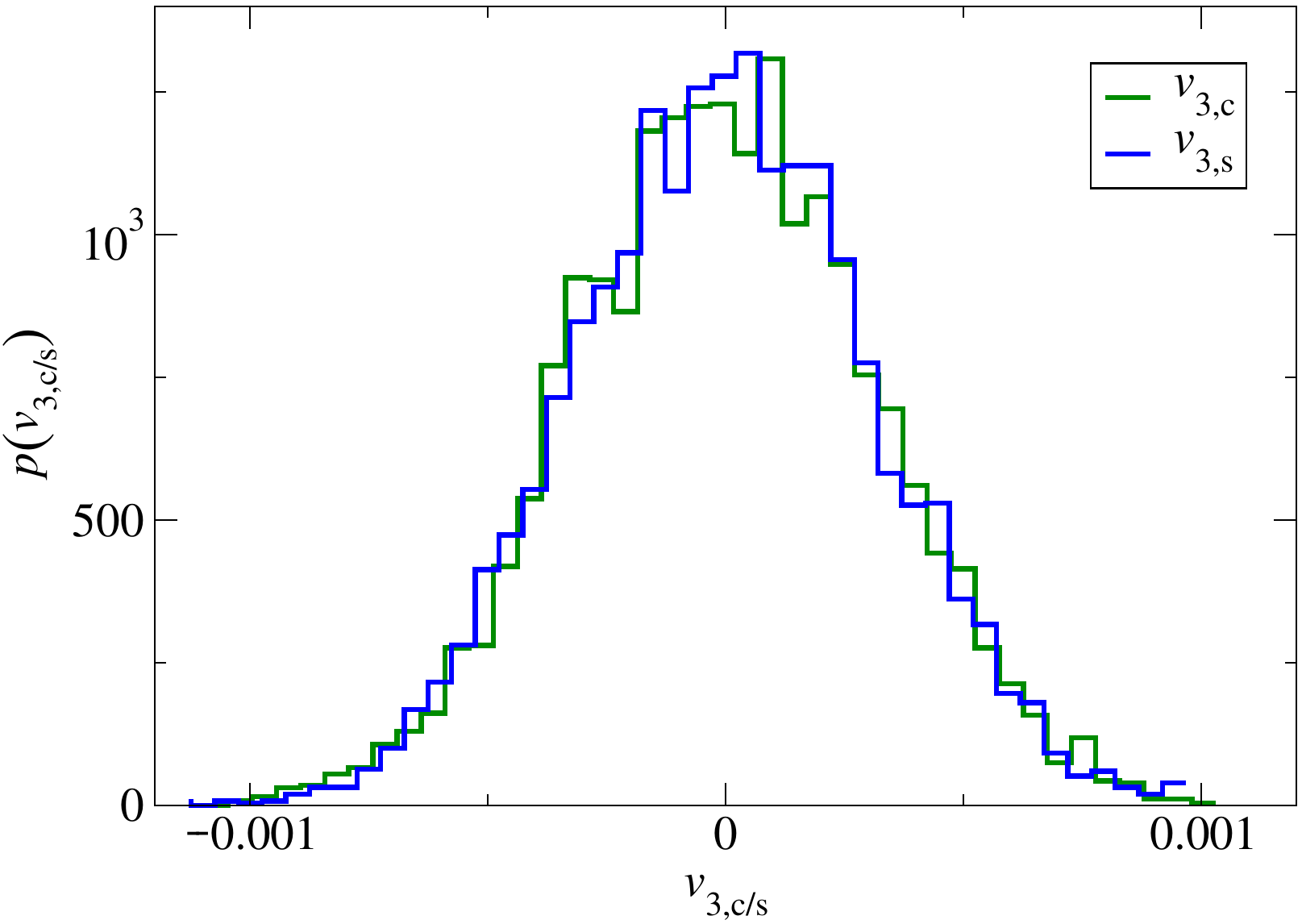}
	\caption{Probability distribution of $v_{2,\mathrm{c}/\mathrm{s}}$ (left) and  $v_{3,\mathrm{c}/\mathrm{s}}$ (right) in the initial state of the transport algorithm computed for 5000 events at $b=6$ fm.}
	\label{fig:initial-vn_hist_b6}
\end{figure*}
\begin{table}[h!]
	\caption{\label{tab:initial-vn_b6}Moments of the probability distributions displayed in Fig.~\ref{fig:initial-vn_hist_b6} for the anisotropic flow harmonics in the initial state computed for 5000 events at $b=6$~fm. 
		The variance $\sigma_v^2$ is divided by the rough estimate $1/2N_{\rm p}$.}
	\begin{ruledtabular}
		\begin{tabular}{cccc}
			& $\sigma_v^2/(2N_{\rm p})^{-1}$ & $|\gamma_1|$ & $\gamma_2$ \\ \hline
			$v_{2,c}$ & $1.022 \pm 0.020$ & $0.149 \pm 0.054$ & $0.000 \pm 0.065$ \\
			$v_{2,s}$ & $1.010 \pm 0.020$ & $0.032 \pm 0.055$ & $0.022 \pm 0.067$ \\
			$v_{3,c}$ & $1.006 \pm 0.020$ & $0.045 \pm 0.052$ & $-0.091 \pm 0.055$ \\
			$v_{3,s}$ & $1.002 \pm 0.020$ & $0.022 \pm 0.056$ & $0.017 \pm 0.069$ \\ 
		\end{tabular}
	\end{ruledtabular}
\end{table}

\begin{figure*}[h]
	\includegraphics*[width=0.495\linewidth]{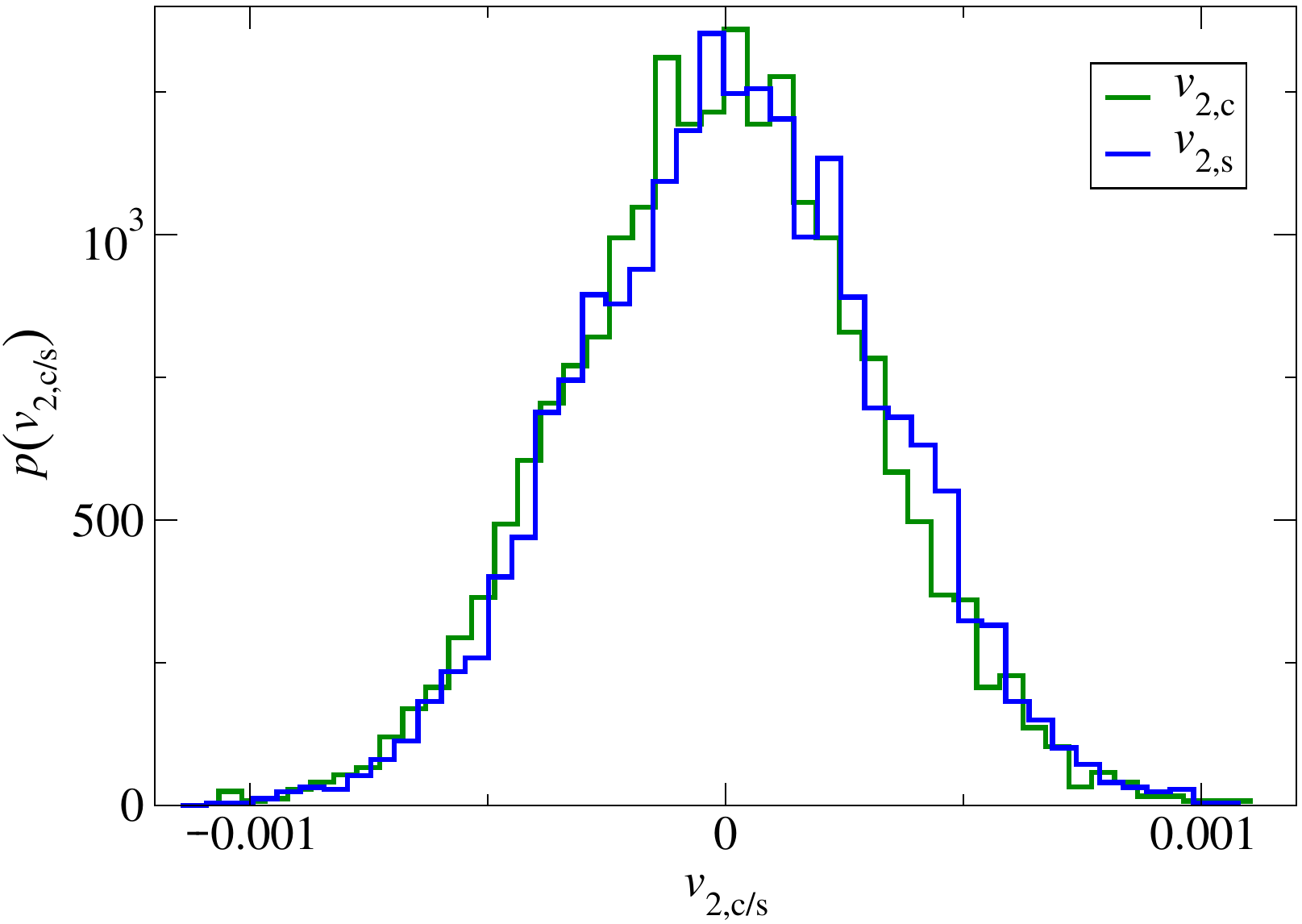}
	\includegraphics*[width=0.495\linewidth]{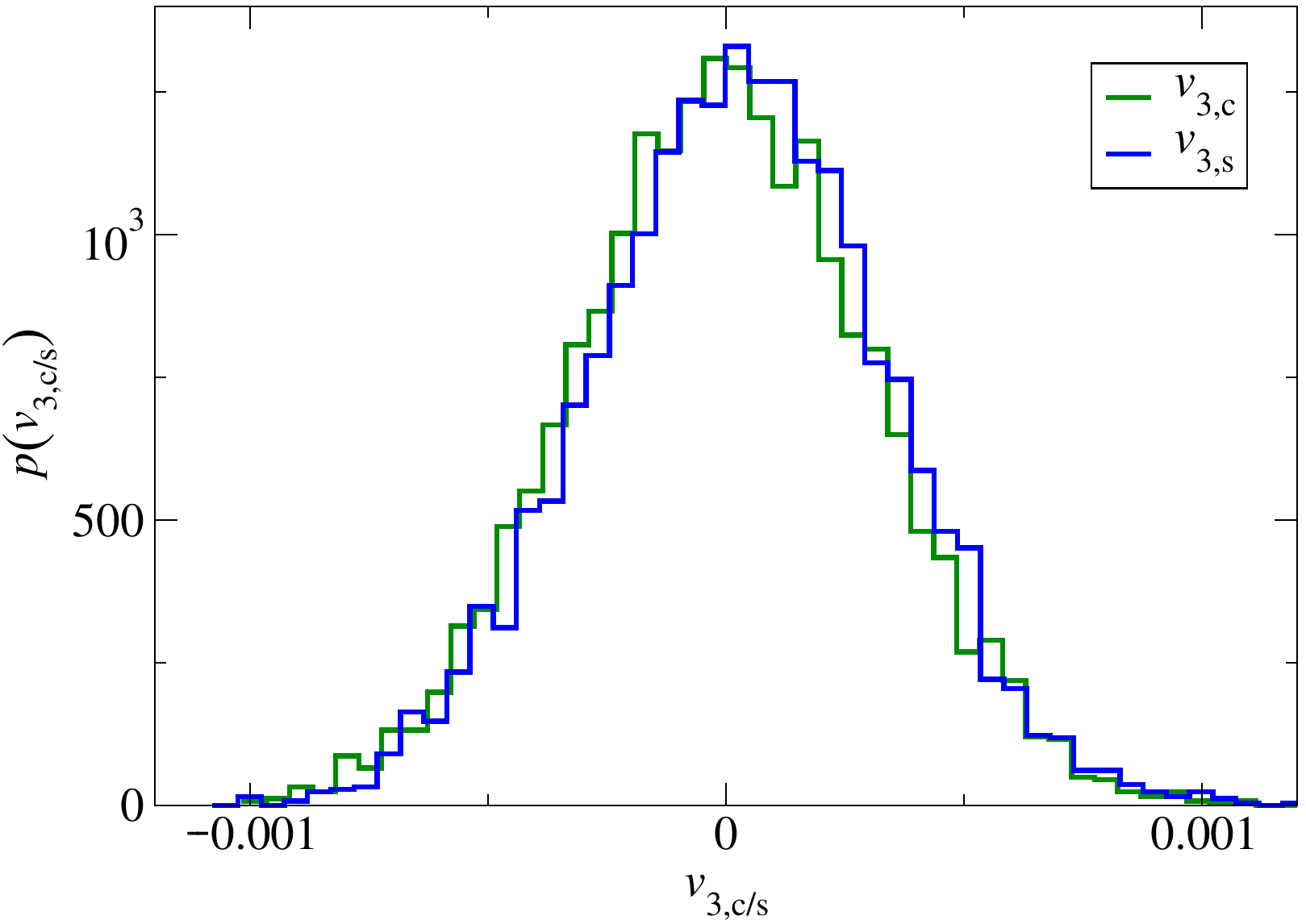}
	\caption{Probability distribution of $v_{2,\mathrm{c}/\mathrm{s}}$ (left) and  $v_{3,\mathrm{c}/\mathrm{s}}$ (right) in the initial state of the transport algorithm computed for 5000 events at $b=9$ fm.}
	\label{fig:initial-vn_hist_b9}
	\vspace{-5mm}
\end{figure*}

\begin{table}[h!]
	\caption{\label{tab:initial-vn_b9}Moments of the probability distributions displayed in Fig.~\ref{fig:initial-vn_hist_b9} for the anisotropic flow harmonics in the initial state computed for 5000 events at $b=9$~fm. 
		The variance $\sigma_v^2$ is divided by the estimate $1/2N_{\rm p}$.}
	\begin{ruledtabular}
		\begin{tabular}{cccc}
			& $\sigma_v^2/(2N_{\rm p})^{-1}$ & $|\gamma_1|$ & $\gamma_2$ \\ \hline
			$v_{2,c}$ & $0.994 \pm 0.020$ & $0.013 \pm 0.055$ & $0.043 \pm 0.066$ \\
			$v_{2,s}$ & $1.002 \pm 0.020$ & $0.110 \pm 0.054$ & $-0.034 \pm 0.071$ \\
			$v_{3,c}$ & $0.994 \pm 0.020$ & $0.109 \pm 0.056$ & $0.021 \pm 0.072$ \\
			$v_{3,s}$ & $0.970 \pm 0.020$ & $0.069 \pm 0.057$ & $0.084 \pm 0.083$ \\
		\end{tabular}
	\end{ruledtabular}\vspace{-3mm}
\end{table}

\section{Fit parameters}
\label{app:fit-params}

In Tables~\ref{tab:fit-params_b0}, \ref{tab:fit-params_b6}, and \ref{tab:fit-params_b9} we gather the parameters of the fit functions to the scatter plots of $v_{n,\rm c/s}$ (in the final state) as function of $\varepsilon_{n,\rm c/s}$. 
Linear fits are used throughout, except for $v_{2,\rm c/s}$ at $b=6$ and 9~fm, where the nonlinear ansatz~\eqref{eq:v2_vs_eps2_nonlin} yields better fits. 

The fit parameters are consistently very similar for cosine and sine harmonics under the same conditions of $b$ and $\langle\textrm{Kn}\rangle^{-1}$.
All fit parameters increase with $\langle\textrm{Kn}\rangle^{-1}$ at a given impact parameter. 
This is of course long known for the linear parts of the fits, but also holds for the nonlinear parameters ${\cal K}_{2,222}$, as recently reported in Ref.~\cite{Kurkela:2020wwb}. 
\begin{table*}[h!]
	\caption{\label{tab:fit-params_b0}Parameters of the linear-fit functions~\eqref{eq:v2_vs_eps2_lin}--\eqref{eq:v3_vs_eps3_lin} for $v_{2,\rm c/s}$ and  $v_{3,\rm c/s}$ at $b=0$~fm.}
	\begin{ruledtabular}
		\begin{tabular}{ccccc}
			$\langle{\rm Kn}\rangle^{-1}$ & $v_{2,\rm c}$  & $v_{2,\rm s}$ & $v_{3,\rm c}$ & $v_{3,\rm s}$ \\
			& ${\cal K}_{2,2}$ & ${\cal K}_{2,2}$ &  ${\cal K}_{3,3}$ & ${\cal K}_{3,3}$ \\ \hline
			0.30 & $0.0455 \pm 0.0001$ & $0.0455 \pm 0.0001$ & 
			$0.0058 \pm 0.0001$ & $0.0058 \pm 0.0001$ \\
			3.01 & $0.1712 \pm 0.0003$ & $0.1706 \pm 0.0003$ & 
			$0.0920 \pm 0.0002$ & $0.0920 \pm 0.0002$ \\
			12.06 & $0.2133 \pm 0.0006$ & $0.2120 \pm 0.0006$ & 
			$0.1490 \pm 0.0006$ & $0.1490 \pm 0.0006$ \\
		\end{tabular}
	\end{ruledtabular}\vspace{-5mm}
\end{table*}

\begin{table*}[h!]
	\caption{\label{tab:fit-params_b6}Parameters of the fit functions for $v_{2,\rm c/s}$ [nonlinear fit Eq.~\eqref{eq:v2_vs_eps2_nonlin}] and  $v_{3,\rm c/s}$ [linear fit Eq.~\eqref{eq:v3_vs_eps3_lin}] at $b=6$~fm.}
	\begin{ruledtabular}
		\begin{tabular}{ccccccc}
			$\langle{\rm Kn}\rangle^{-1}$ & \multicolumn{2}{c}{$v_{2,\rm c}$}  & \multicolumn{2}{c}{$v_{2,\rm s}$} & $v_{3,\rm c}$ & $v_{3,\rm s}$ \\
			& ${\cal K}_{2,2}$ & ${\cal K}_{2,222}$ & ${\cal K}_{2,2}$ & ${\cal K}_{2,222}$ &  ${\cal K}_{3,3}$ & ${\cal K}_{3,3}$ \\ \hline
			0.34 & $0.0589 \pm 0.0001$ & $0.0018 \pm 0.0015$ & $0.0565 \pm 0.0002$ & 
			$0.0171 \pm 0.0065$ & $0.00791 \pm 0.00004$ & $0.00782 \pm 0.00004$ \\
			3.43 & $0.1924 \pm 0.0003$ & $0.0182 \pm 0.0043$ & $0.1900 \pm 0.0005$ & 
			$0.0458 \pm 0.0203$ & $0.1069 \pm 0.0002$ & $0.1062 \pm 0.0002$ \\
			13.73 & $0.2382 \pm 0.0005$ & $0.0235 \pm 0.0082$ & $0.2379 \pm 0.0009$ & 
			$0.0539 \pm 0.0391$ & $0.1770 \pm 0.0005$ & $0.1756 \pm 0.0005$ \\
		\end{tabular}
	\end{ruledtabular}
\end{table*}

\begin{table*}[h!]
	\caption{\label{tab:fit-params_b9}Parameters of the fit functions for $v_{2,\rm c/s}$ [nonlinear fit Eq.~\eqref{eq:v2_vs_eps2_nonlin}] and  $v_{3,\rm c/s}$ [linear fit Eq.~\eqref{eq:v3_vs_eps3_lin}] at $b=9$~fm.}
	\begin{ruledtabular}
		\begin{tabular}{ccccccc}
			$\langle{\rm Kn}\rangle^{-1}$ & \multicolumn{2}{c}{$v_{2,\rm c}$}  & \multicolumn{2}{c}{$v_{2,\rm s}$} & $v_{3,\rm c}$ & $v_{3,\rm s}$ \\
			& ${\cal K}_{2,2}$ & ${\cal K}_{2,222}$ & ${\cal K}_{2,2}$ & ${\cal K}_{2,222}$ &  ${\cal K}_{3,3}$ & ${\cal K}_{3,3}$ \\ \hline
			0.41 & $0.0662 \pm 0.0001$ & $0.0161 \pm 0.0007$ & $0.0646 \pm 0.0002$ & 
			$0.0081 \pm 0.0031$ & $0.00989 \pm 0.00003$ & $0.00970 \pm 0.00003$ \\
			4.10 & $0.1981 \pm 0.0002$ & $0.0425 \pm 0.0017$ & $0.1976 \pm 0.0004$ & 
			$0.0400 \pm 0.0086$ & $0.1106 \pm 0.0002$ & $0.1091 \pm 0.0003$ \\
			16.42 & $0.2414 \pm 0.0004$ & $0.0456 \pm 0.0029$ & $0.2422 \pm 0.0008$ & 
			$0.0520 \pm 0.0150$ & $0.1697 \pm 0.0005$ & $0.1670 \pm 0.0005$ \\
		\end{tabular}
	\end{ruledtabular}
\end{table*}

\section{Complex eccentricity and elliptic flow at $b=6$~fm}
\label{app:complex_eps2,v2}

In this Appendix, we show for $10^4$ events at $b=6$~fm the dispersion of the complex eccentricity $\varepsilon_{2\,}{\rm e}^{2{\rm i}\Phi_2}$ in the initial state (Fig.~\ref{fig:complex-eps2_binned_b6} --- remember that the impact parameter lies at $\Phi_2=0$) and of the final state elliptic flow $v_{2\,}{\rm e}^{2{\rm i}\Psi_2}$ for the smallest and largest values of the inverse Knudsen number (Fig.~\ref{fig:complex-v2_binned_b6}).
We also indicate the five bins in $\varepsilon_{2,\mathrm{c}}$ for those events (see also Fig.~\ref{fig:Scatter_b6}). 
In particular, one sees that they do not map onto similar bins in $v_{2,c} = {\rm Re}\big(v_{2\,}{\rm e}^{2{\rm i}\Psi_2}\big)$, due to the fluctuation of the response. 
Figure~\ref{fig:complex-eps2_binned_b6} also shows that the events with ``outlying'' $\varepsilon_{2,\mathrm{c}}$ that we remove from our statistical analysis of $v_{2,\rm c}$ do not have at the same time an ``outlying'' $\varepsilon_{2,\mathrm{s}}$, and thus are not the same as are left aside in the study of $v_{2,\rm s}$.

\begin{figure*}[ht!]
	\includegraphics*[width=0.495\linewidth]{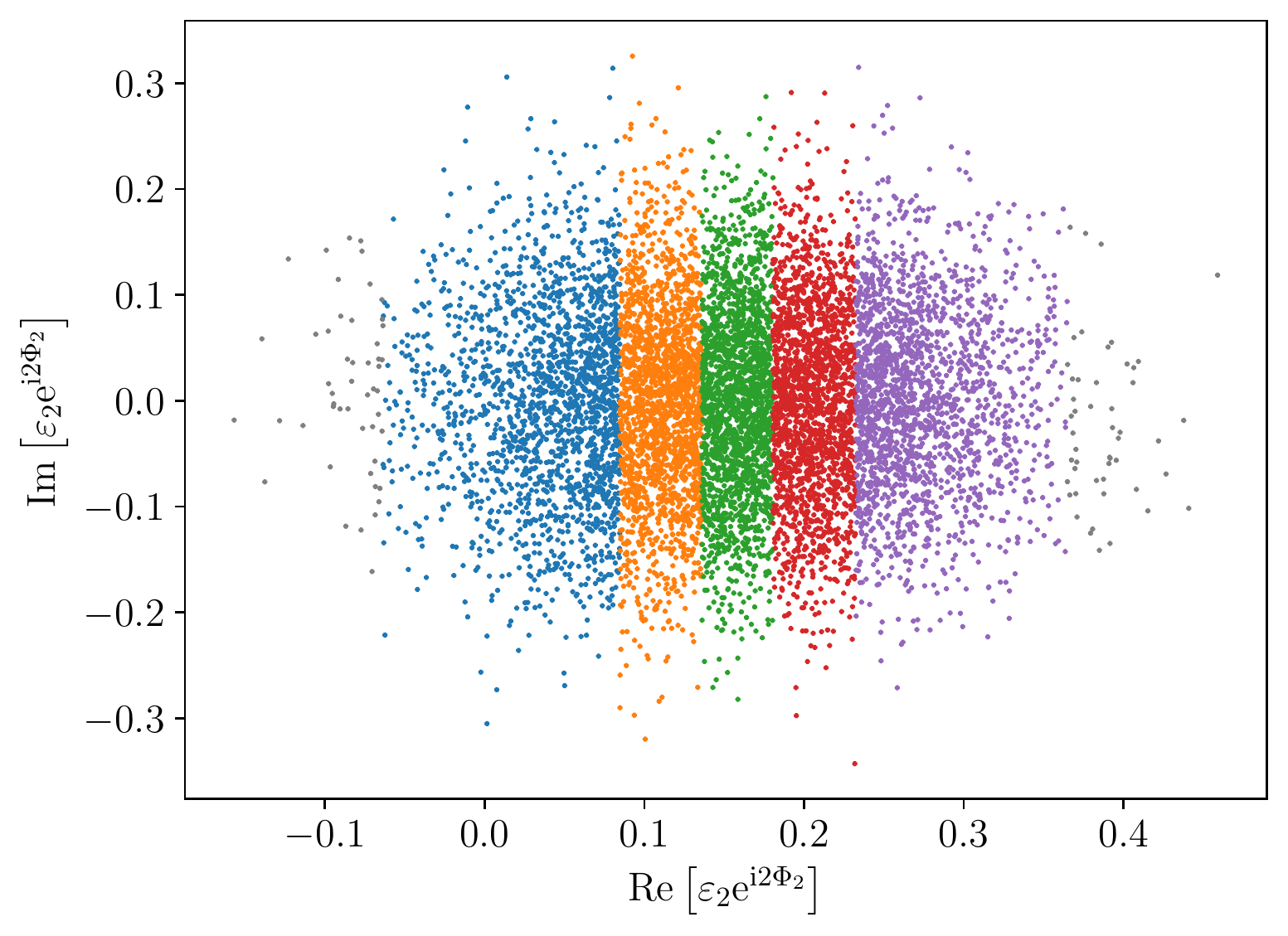}
	\caption{Scatter plot of the initial complex eccentricity $\varepsilon_{2\,}{\rm e}^{2{\rm i}\Phi_2} \equiv \varepsilon_{2,c} + {\rm i}_{}\varepsilon_{2,s}$ for $10^4$ events at $b=6$ fm.
	The bins are the same as in Fig.~\ref{fig:Scatter_b6}.}
	\label{fig:complex-eps2_binned_b6}
\end{figure*}
\begin{figure*}[ht!]
	\includegraphics*[width=0.495\linewidth]{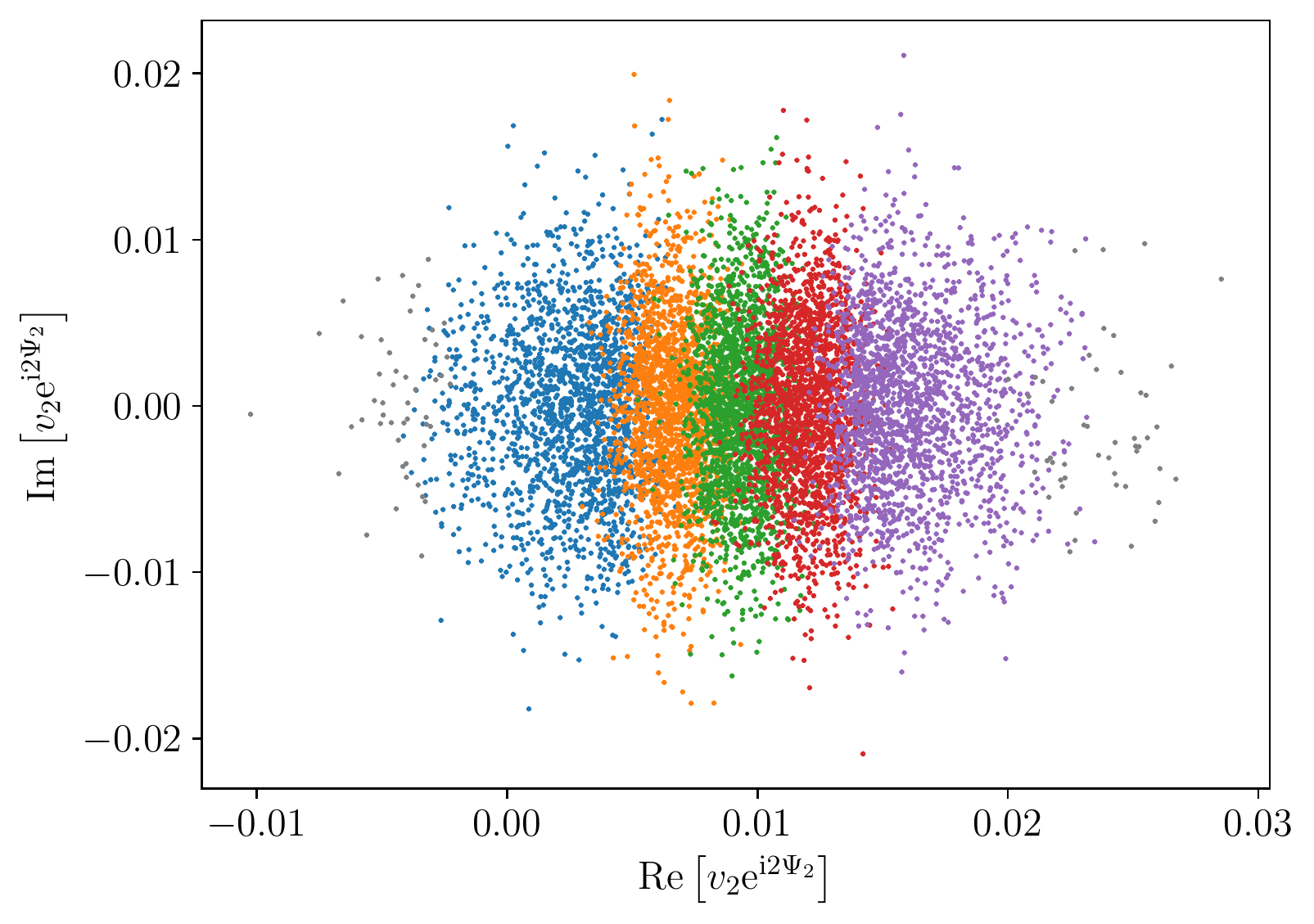}
	\includegraphics*[width=0.495\linewidth]{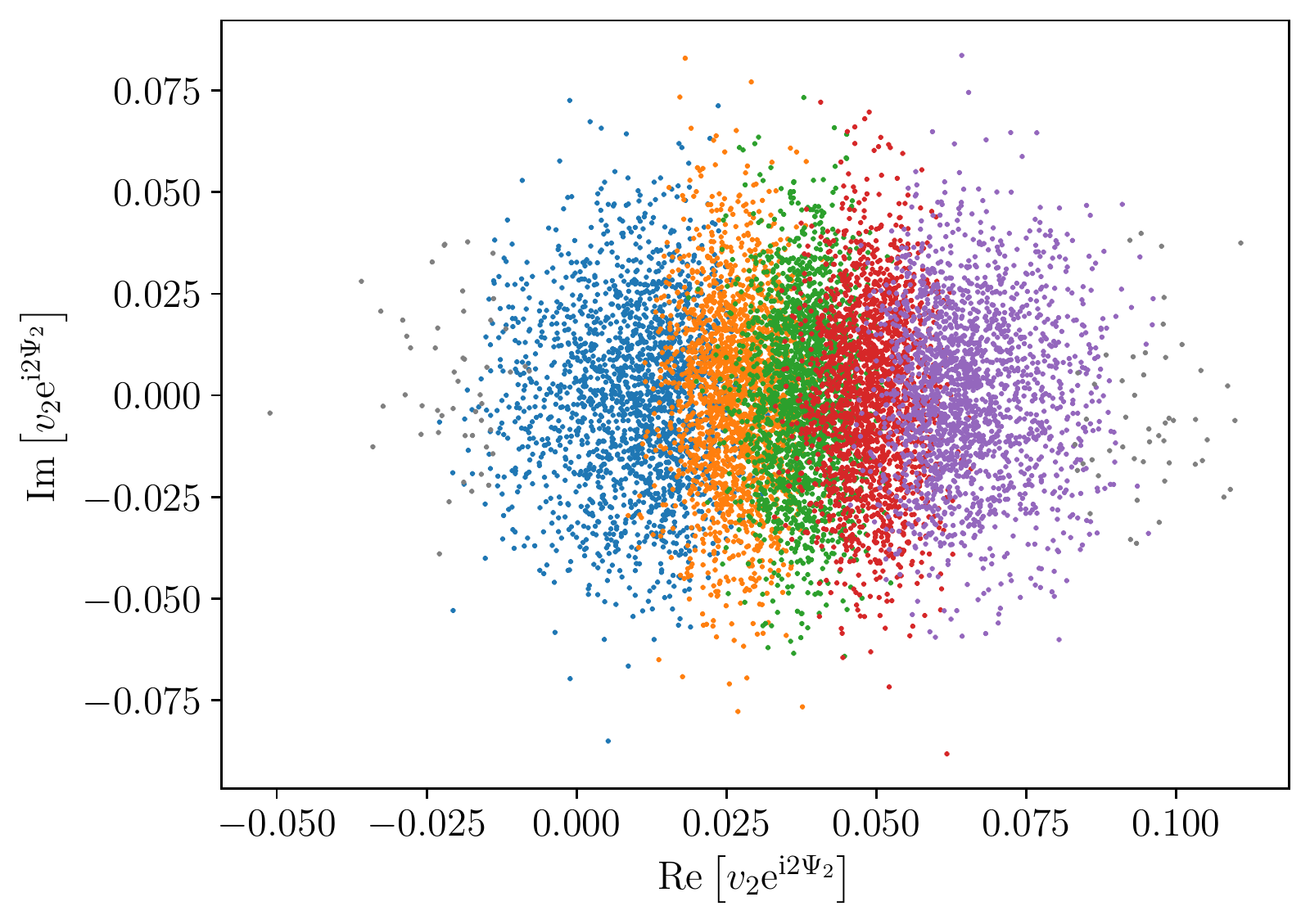}
	\caption{Scatter plot of the final complex elliptic flow $v_{2\,}{\rm e}^{2{\rm i}\Psi_2} \equiv v_{2,c} + {\rm i}_{}v_{2,s}$ for events at $b=6$~fm with $\langle \textrm{Kn}\rangle^{-1}=0.34$ (left) and $\langle \textrm{Kn}\rangle^{-1}=13.74$ (right).
	The bins are those in $\varepsilon_{2,\mathrm{c}}$ as in Fig.~\ref{fig:Scatter_b6} and \ref{fig:complex-eps2_binned_b6}.}
	\label{fig:complex-v2_binned_b6}
\end{figure*}

\section{Averaged moments of the anisotropic flow harmonics}
\label{app:moments_b0-9}

In this Appendix, we list the moments of the conditional probability distributions $p_{v|\varepsilon}(v_{n,\rm c/s}|\varepsilon_{n,\rm c/s})$ about the mean value $\bar{v}_{n,\rm c/s}(\varepsilon_{n,\rm c/s})$ for our sets of events at $b=0$ (Tables~\ref{tab:final-v2_b0} and \ref{tab:final-v3_b0}) and 9~fm (Tables~\ref{tab:final-v2_b9} and \ref{tab:final-v3_b9}), complementing the results at 6~fm presented in Sec.~\ref{ss:results_moments}.

\begin{table*}[h!]
	\caption{\label{tab:final-v2_b0}Moments of the conditional probability distributions $p_{v|\varepsilon}(v_{2,\rm c/s}|\varepsilon_{2,\rm c/s})$ of the final-state flow harmonics computed for $10^4$ events at $b=0$~fm. 
		The variance $\sigma_v^2$ is divided by the square ${\cal K}_{2,2}^2$ of the linear coefficient in relation~\eqref{eq:v2_vs_eps2_lin}.}
	\begin{ruledtabular}
		\begin{tabular}{ccccccc}
			$\langle{\rm Kn}\rangle^{-1}$ & \multicolumn{3}{c}{$v_{2,\rm c}$}  & \multicolumn{3}{c}{$v_{2,\rm s}$} \\
			& $10^4\,\sigma_v^2/{\cal K}_{2,2}^2$ & $|\gamma_1|$ & $\gamma_2$ & 
			$10^4\,\sigma_v^2/{\cal K}_{2,2}^2$ & $|\gamma_1|$ & $\gamma_2$\\ \hline
			0.30 & $1.482 \pm 0.049$ & $0.112 \pm 0.094$ & $0.161 \pm 0.125$ & 
			$1.489 \pm 0.049$ & $0.295 \pm 0.089$ & $0.091 \pm 0.111$ \\
			3.01 & $0.905 \pm 0.030$ & $0.309 \pm 0.092$ & $0.219 \pm 0.124$ & 
			$0.908 \pm 0.031$ & $0.416 \pm 0.095$ & $0.252 \pm 0.140$ \\
			12.06 & $2.212 \pm 0.071$ & $0.372 \pm 0.084$ & $0.041 \pm 0.100$ & 
			$2.183 \pm 0.070$ & $0.081 \pm 0.090$ & $0.046 \pm 0.120$ \\
		\end{tabular}
	\end{ruledtabular}
\end{table*}

\begin{table*}[h!]
	\caption{\label{tab:final-v3_b0}Moments of the conditional probability distributions $p_{v|\varepsilon}(v_{3,\rm c/s}|\varepsilon_{3,\rm c/s})$ of the final-state flow harmonics computed for $10^4$ events at $b=0$~fm. 
		The variance $\sigma_v^2$ is divided by the square ${\cal K}_{3,3}^2$ of the linear coefficient in relation~\eqref{eq:v3_vs_eps3_lin}.}
	\begin{ruledtabular}
		\begin{tabular}{ccccccc}
			$\langle{\rm Kn}\rangle^{-1}$ & \multicolumn{3}{c}{$v_{3,\rm c}$}  & \multicolumn{3}{c}{$v_{3,\rm s}$} \\
			& $10^4\,\sigma_v^2/{\cal K}_{3,3}^2$ & $|\gamma_1|$ & $\gamma_2$ & 
			$10^4\,\sigma_v^2/{\cal K}_{3,3}^2$ & $|\gamma_1|$ & $\gamma_2$\\ \hline
			0.30 & $32.27 \pm 1.03$ & $0.313 \pm 0.083$ & $-0.024 \pm 0.095$ & 
			$37.58 \pm 1.15$ & $1.053 \pm 0.054$ & $-0.144 \pm 0.084$ \\
			3.01 & $1.663 \pm 0.044$ & $0.074 \pm 0.093$ & $0.052 \pm 0.128$ & 
			$0.183 \pm 0.058$ & $0.875 \pm 0.065$ & $-0.080 \pm 0.094$ \\
			12.06 & $5.488 \pm 0.175$ & $0.061 \pm 0.088$ & $-0.031 \pm 0.109$ & 
			$0.553 \pm 0.173$ & $0.407 \pm 0.080$ & $-0.064 \pm 0.096$ \\
		\end{tabular}
	\end{ruledtabular}
\end{table*}

\begin{table*}[h!]
	\caption{\label{tab:final-v2_b9}Moments of the conditional probability distributions $p_{v|\varepsilon}(v_{2,\rm c/s}|\varepsilon_{2,\rm c/s})$ of the final-state flow harmonics computed for $10^4$ events at $b=9$~fm. 
		The variance $\sigma_v^2$ is divided by the square ${\cal K}_{2,2}^2$ of the linear coefficient in relation~\eqref{eq:v2_vs_eps2_nonlin}.}
	\begin{ruledtabular}
		\begin{tabular}{ccccccc}
			$\langle{\rm Kn}\rangle^{-1}$ & \multicolumn{3}{c}{$v_{2,\rm c}$}  & \multicolumn{3}{c}{$v_{2,\rm s}$} \\
			& $10^4\,\sigma_v^2/{\cal K}_{2,2}^2$ & $|\gamma_1|$ & $\gamma_2$ & 
			$10^4\,\sigma_v^2/{\cal K}_{2,2}^2$ & $|\gamma_1|$ & $\gamma_2$\\ \hline
			0.41 & $3.559 \pm 0.149$ & $0.606 \pm 0.127$ & $0.551 \pm 0.269$ & 
			$2.688 \pm 0.133$ & $0.383 \pm 0.196$ & $1.334 \pm 0.440$ \\
			4.10 & $2.376 \pm 0.101$ & $0.683 \pm 0.130$ & $1.147 \pm 0.403$ & 
			$2.423 \pm 0.104$ & $0.381 \pm 0.157$ & $0.995 \pm 0.372$ \\
			16.42 & $4.837 \pm 0.180$ & $0.404 \pm 0.112$ & $0.733 \pm 0.267$ & 
			$4.940 \pm 0.178$ & $0.219 \pm 0.113$ & $0.379 \pm 0.187$ \\
		\end{tabular}
	\end{ruledtabular}
\end{table*}

\begin{table*}[h!]
	\caption{\label{tab:final-v3_b9}Moments of the conditional probability distributions $p_{v|\varepsilon}(v_{3,\rm c/s}|\varepsilon_{3,\rm c/s})$ of the final-state flow harmonics computed for $10^4$ events at $b=9$~fm. 
		The variance $\sigma_v^2$ is divided by the square ${\cal K}_{3,3}^2$ of the linear coefficient in relation~\eqref{eq:v3_vs_eps3_lin}.}
	\begin{ruledtabular}
		\begin{tabular}{ccccccc}
			$\langle{\rm Kn}\rangle^{-1}$ & \multicolumn{3}{c}{$v_{3,\rm c}$}  & \multicolumn{3}{c}{$v_{3,\rm s}$} \\
			& $10^3\,\sigma_v^2/{\cal K}_{3,3}^2$ & $|\gamma_1|$ & $\gamma_2$ & 
			$10^3\,\sigma_v^2/{\cal K}_{3,3}^2$ & $|\gamma_1|$ & $\gamma_2$\\ \hline
			0.41 & $2.190 \pm 0.080$ & $0.138 \pm 0.116$ & $0.608 \pm 0.251$ & 
			$2.601 \pm 0.089$ & $0.942 \pm 0.078$ & $0.171 \pm 0.125$ \\
			4.10 & $0.823 \pm 0.034$ & $0.443 \pm 0.138$ & $1.262 \pm 0.351$ & 
			$1.000 \pm 0.038$ & $0.386 \pm 0.117$ & $0.548 \pm 0.195$ \\
			16.42 & $1.304 \pm 0.045$ & $0.209 \pm 0.098$ & $0.267 \pm 0.159$ & 
			$1.499 \pm 0.050$ & $0.221 \pm 0.100$ & $0.264 \pm 0.180$ \\
		\end{tabular}
	\end{ruledtabular}
\end{table*}

\section{Event-plane angle distributions}
\label{appendix:rp_dist}

\begin{figure*}[ht!]
	\centering
	\includegraphics*[width=0.495\linewidth]{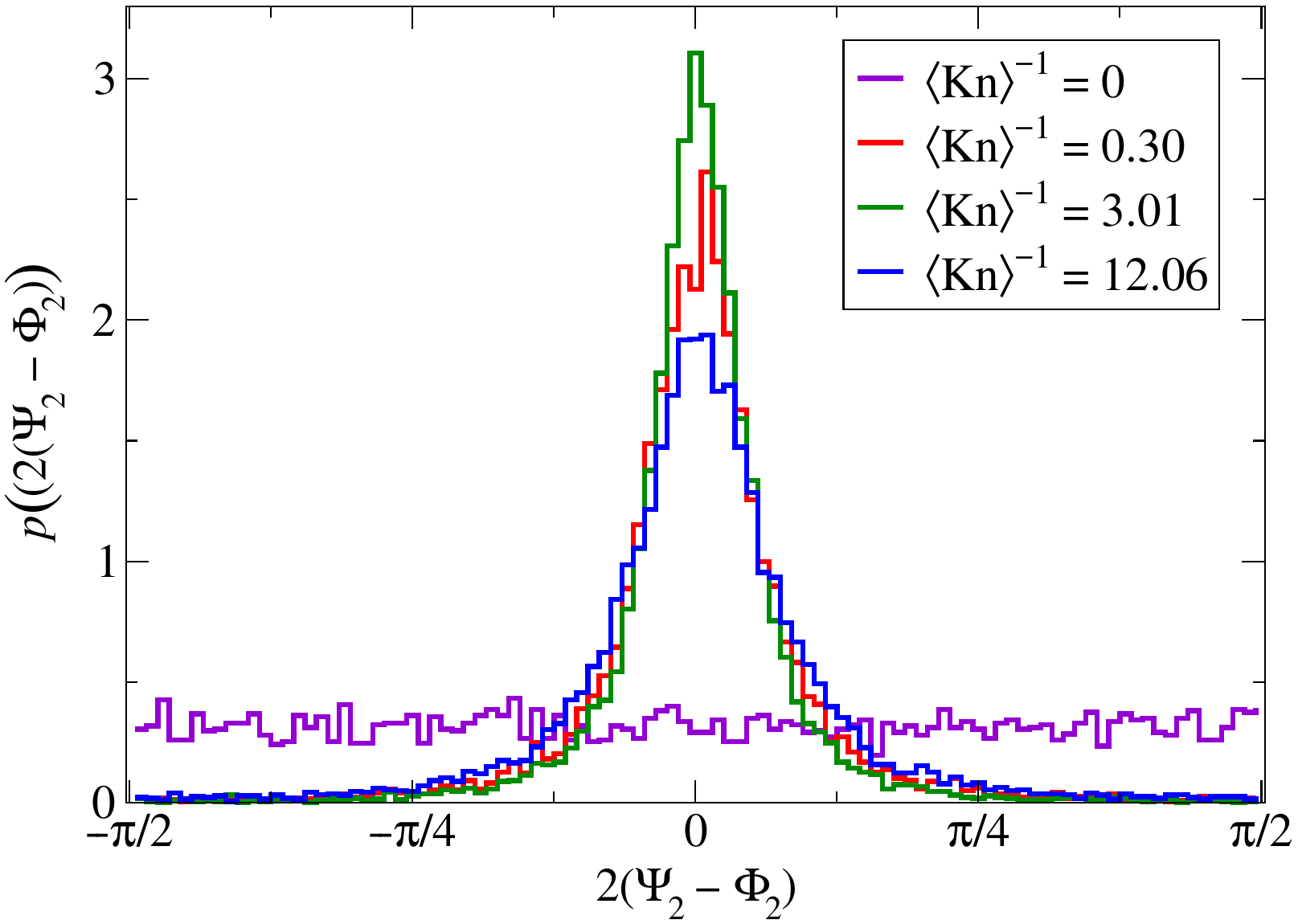}
	\includegraphics[width=0.495\linewidth]{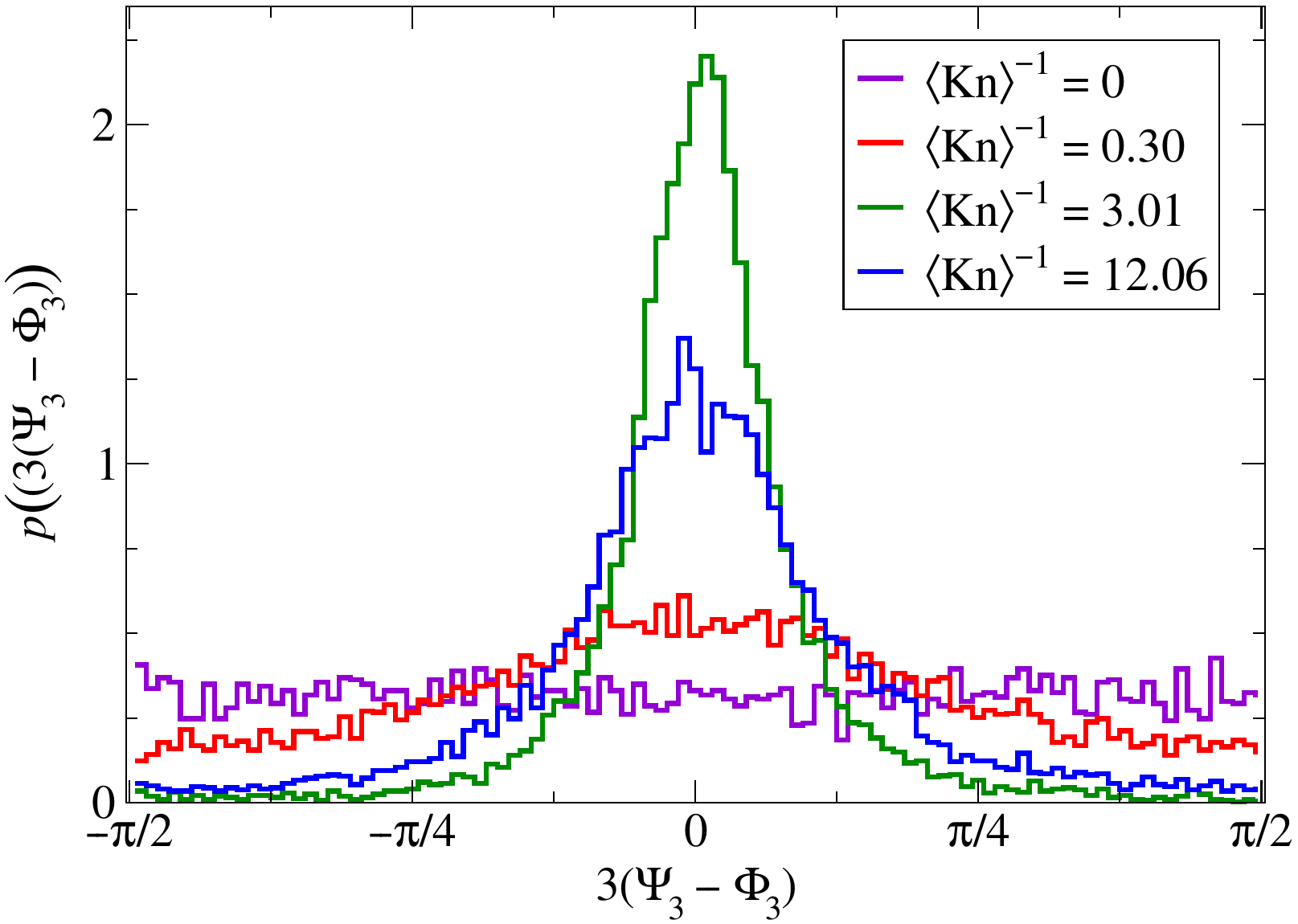}
	\caption{Probability distribution of $p(2(\Psi_2-\Phi_2))$ (left) and $p(3(\Psi_3-\Phi_3))$ (right) at $b=0$~fm for different values of $\langle \textrm{Kn}\rangle^{-1}$.}
	\label{fig:b0_angles}
\end{figure*}

\begin{figure*}[ht!]
	\centering
	\includegraphics*[width=0.495\linewidth]{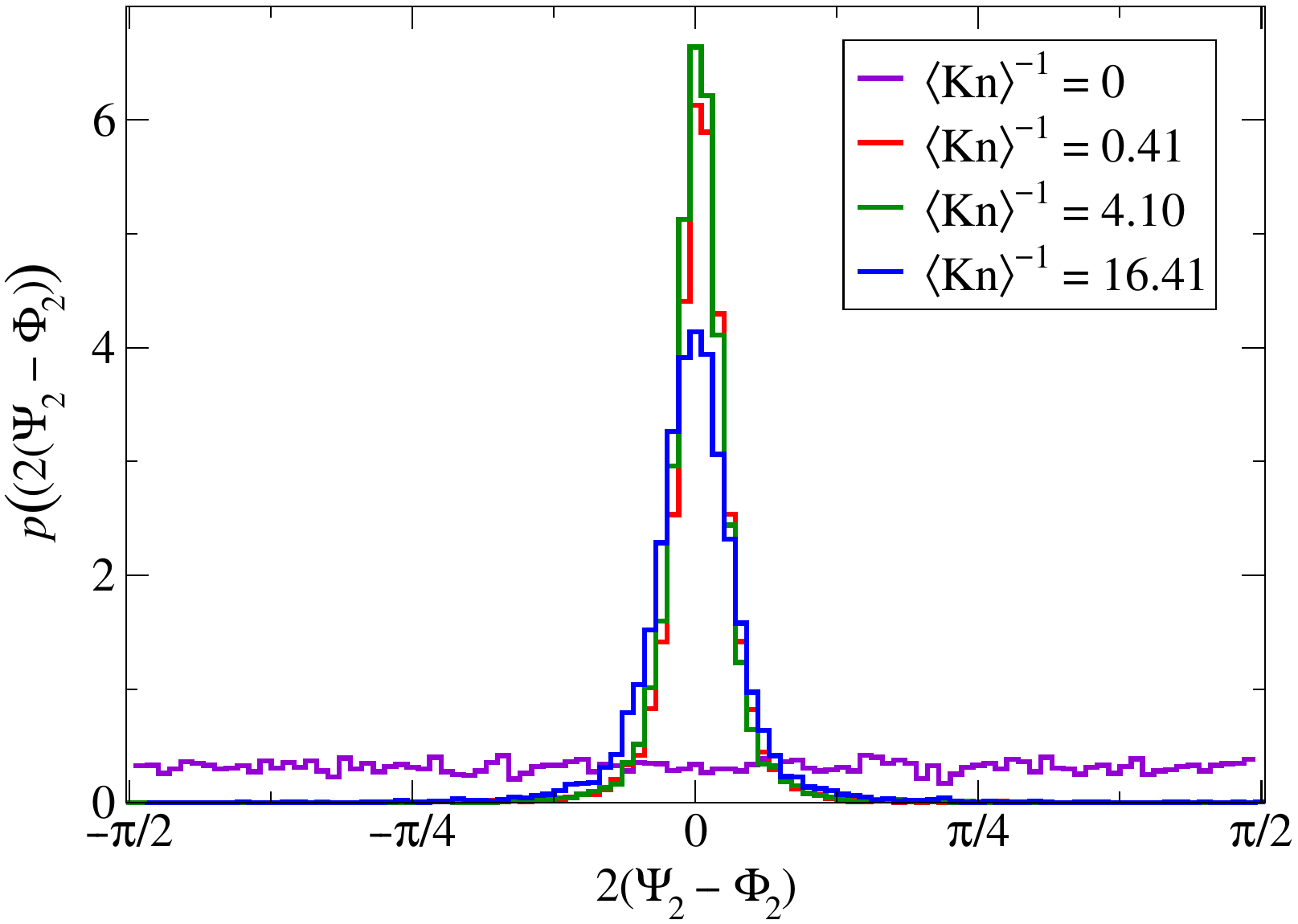}
	\includegraphics[width=0.495\linewidth]{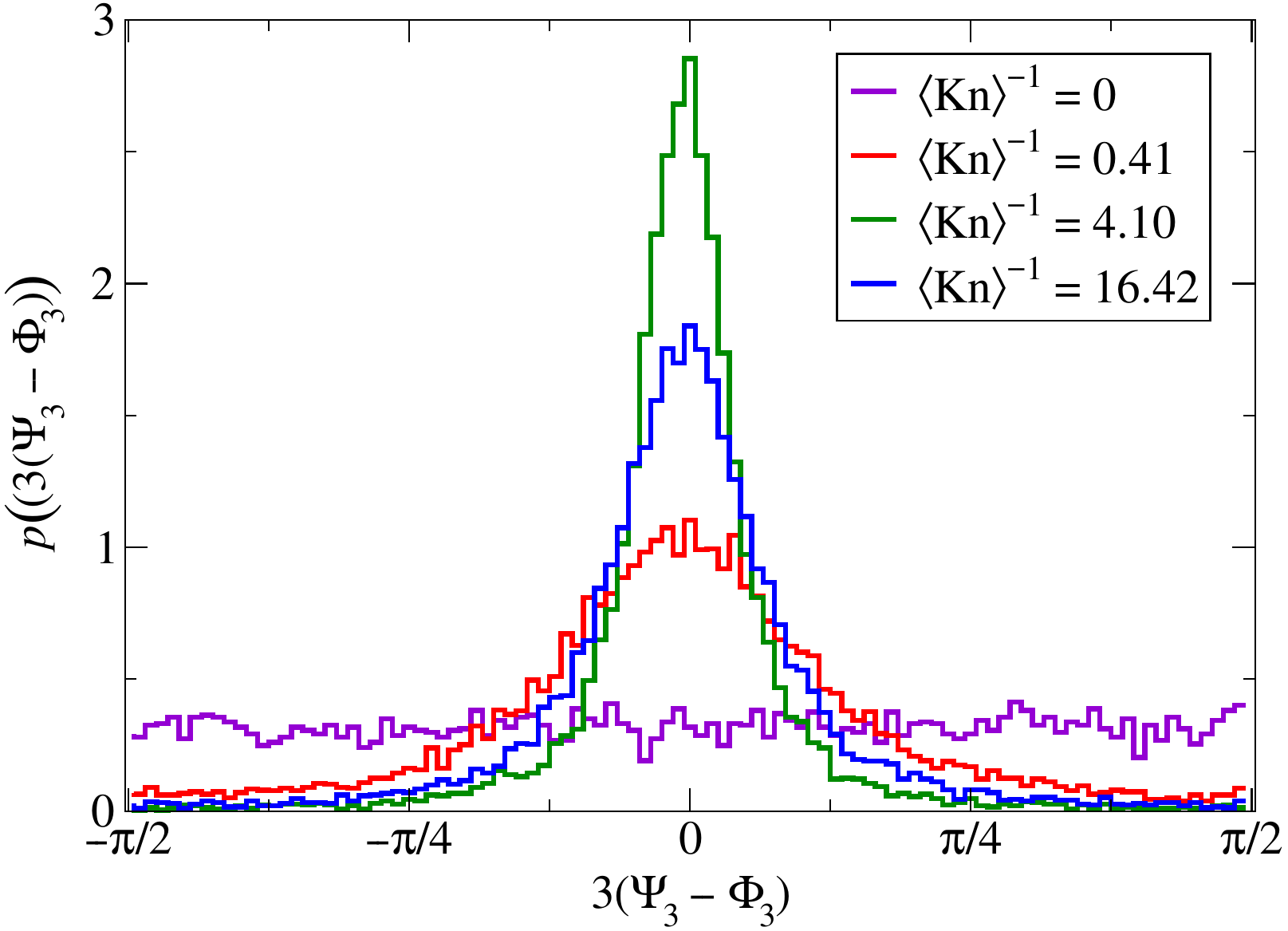}
	\caption{Probability distribution of $p(2(\Psi_2-\Phi_2))$ (left) and $p(3(\Psi_3-\Phi_3))$ (right) at $b=9$~fm for different values of $\langle \textrm{Kn}\rangle^{-1}$.}
	\label{fig:b9_angles}
\end{figure*}

In this Appendix, we show the distributions of each event-plane angle $\Psi_n$ (in the final state) about the corresponding participant-plane orientation $\Phi_n$ (in the initial state) for our sets of events at $b=0$ (Fig.~\ref{fig:b0_angles}) and 9~fm (Fig.~\ref{fig:b9_angles}).

\end{widetext}


\begin{thebibliography}{99}

\bibitem{Heinz:2013th}
U.~Heinz and R.~Snellings,
``Collective flow and viscosity in relativistic heavy-ion collisions,''
Ann. Rev. Nucl. Part. Sci. {\bf 63} (2013) 123
[arXiv:1301.2826 [nucl-th]].

\bibitem{Luzum:2013yya}
M.~Luzum and H.~Petersen,
``Initial State Fluctuations and Final State Correlations in Relativistic Heavy-Ion Collisions,''
J. Phys. G {\bf 41} (2014) 063102
[arXiv:1312.5503 [nucl-th]].

\bibitem{Bhalerao:2020ulk}
R.~S.~Bhalerao,
``Collectivity in large and small systems formed in ultrarelativistic collisions,''
[arXiv:2009.09586 [nucl-th]].

\bibitem{Alver:2010gr}
B.~Alver and G.~Roland,
``Collision geometry fluctuations and triangular flow in heavy-ion collisions,''
Phys.\ Rev.\ C {\bf 81} (2010) 054905 
[arXiv:1003.0194 [nucl-th]]
[Erratum: Phys.\ Rev.\ C {\bf 82} (2010) 039903].

\bibitem{Teaney:2010vd}
D.~Teaney and L.~Yan,
``Triangularity and Dipole Asymmetry in Heavy Ion Collisions,''
Phys.\ Rev.\ C {\bf 83} (2011) 064904
[arXiv:1010.1876 [nucl-th]].

\bibitem{Gardim:2011xv}
F.~G.~Gardim, F.~Grassi, M.~Luzum and J.-Y.~Ollitrault,
``Mapping the hydrodynamic response to the initial geometry in heavy-ion collisions,''
Phys.\ Rev.\ C {\bf 85} (2012) 024908  
[arXiv:1111.6538 [nucl-th]].

\bibitem{Voloshin:1994mz}
S.~Voloshin and Y.~Zhang,
``Flow study in relativistic nuclear collisions by Fourier expansion of azimuthal particle distributions,''
Z. Phys. C {\bf 70} (1996) 665
[arXiv:hep-ph/9407282 [hep-ph]].
	
\bibitem{Ollitrault:1992bk}
J.-Y.~Ollitrault,
``Anisotropy as a signature of transverse collective flow,''
Phys.\ Rev.\ D {\bf 46} (1992) 229.
	
\bibitem{Borghini:2010hy}
N.~Borghini and C.~Gombeaud,
``Anisotropic flow far from equilibrium,''
Eur. Phys. J. C \textbf{71} (2011) 1612
[arXiv:1012.0899 [nucl-th]].

\bibitem{Niemi:2012aj}
H.~Niemi, G.~S.~Denicol, H.~Holopainen and P.~Huovinen,
``Event-by-event distributions of azimuthal asymmetries in ultrarelativistic heavy-ion collisions,''
Phys.\ Rev.\ C {\bf 87} (2013) 054901
[arXiv:1212.1008 [nucl-th]].
	
\bibitem{Plumari:2015cfa}
S.~Plumari, G.~L.~Guardo, F.~Scardina and V.~Greco,
``Initial state fluctuations from mid-peripheral to ultra-central collisions in a event-by-event transport approach,''
Phys.\ Rev.\ C {\bf 92} (2015) 054902
[arXiv:1507.05540 [hep-ph]].

\bibitem{Noronha-Hostler:2015dbi}
J.~Noronha-Hostler, L.~Yan, F.~G.~Gardim and J.-Y.~Ollitrault,
``Linear and cubic response to the initial eccentricity in heavy-ion collisions,''
Phys.\ Rev.\ C {\bf 93} (2016) 014909
[arXiv:1511.03896 [nucl-th]].

\bibitem{Miller:2003kd}
M.~Miller and R.~Snellings,
``Eccentricity fluctuations and its possible effect on elliptic flow measurements,''
[arXiv:nucl-ex/0312008 [nucl-ex]].

\bibitem{Miller:2007ri}
M.~L.~Miller, K.~Reygers, S.~J.~Sanders and P.~Steinberg,
``Glauber modeling in high energy nuclear collisions,''
Ann. Rev. Nucl. Part. Sci. {\bf 57} (2007) 205
[arXiv:nucl-ex/0701025 [nucl-ex]].

\bibitem{Yan:2014nsa}
L.~Yan, J.~Y.~Ollitrault and A.~M.~Poskanzer,
``Azimuthal Anisotropy Distributions in High-Energy Collisions,''
Phys. Lett. B {\bf 742} (2015) 290
[arXiv:1408.0921 [nucl-th]].

\bibitem{Aad:2013xma}
G.~Aad \textit{et al.} [ATLAS Collaboration],
``Measurement of the distributions of event-by-event flow harmonics in lead-lead collisions at $\sqrt{{\mathrm{s}}_{\mathrm{NN}}}$ = 2.76 TeV with the ATLAS detector at the LHC,''
JHEP {\bf11} (2013) 183
[arXiv:1305.2942 [hep-ex]].

\bibitem{ALICE:2016kpq}
J.~Adam \textit{et al.} [ALICE Collaboration],
``Correlated event-by-event fluctuations of flow harmonics in Pb-Pb collisions at $\sqrt{s_{_{\rm NN}}}=2.76$ TeV,''
Phys. Rev. Lett. {\bf 117} (2016) 182301
[arXiv:1604.07663 [nucl-ex]].

\bibitem{Sirunyan:2017fts}
A.~M.~Sirunyan \textit{et al.} [CMS Collaboration],
``Non-Gaussian elliptic-flow fluctuations in PbPb collisions at $\sqrt{\smash[b]{s_{_\text{NN}}}} = 5.02$ TeV,''
Phys. Lett. B {\bf 789} (2019) 643
[arXiv:1711.05594 [nucl-ex]].

\bibitem{Aaboud:2019sma}
M.~Aaboud \textit{et al.} [ATLAS Collaboration],
``Fluctuations of anisotropic flow in Pb+Pb collisions at $\sqrt{{\mathrm{s}}_{\mathrm{NN}}}$ = 5.02 TeV with the ATLAS detector,''
JHEP {\bf 01} (2020), 051
[arXiv:1904.04808 [nucl-ex]].

\bibitem{Petersen:2010cw}
H.~Petersen, G.~Y.~Qin, S.~A.~Bass and B.~Muller,
``Triangular flow in event-by-event ideal hydrodynamics in Au+Au collisions at $\sqrt{s_{\rm NN}}=200A$ GeV,''
Phys. Rev. C {\bf 82} (2010) 041901
[arXiv:1008.0625 [nucl-th]].

\bibitem{Qiu:2011iv}
Z.~Qiu and U.~W.~Heinz,
``Event-by-event shape and flow fluctuations of relativistic heavy-ion collision fireballs,''
Phys. Rev. C {\bf 84} (2011) 024911
[arXiv:1104.0650 [nucl-th]].

\bibitem{Gale:2012rq}
C.~Gale, S.~Jeon, B.~Schenke, P.~Tribedy and R.~Venugopalan,
``Event-by-event anisotropic flow in heavy-ion collisions from combined Yang--Mills and viscous fluid dynamics,''
Phys. Rev. Lett. {\bf 110} (2013) 012302
[arXiv:1209.6330 [nucl-th]].

\bibitem{Teaney:2013dta}
D.~Teaney and L.~Yan,
``Event-plane correlations and hydrodynamic simulations of heavy ion collisions,''
Phys. Rev. C {\bf 90} (2014) 024902
[arXiv:1312.3689 [nucl-th]].

\bibitem{Ma:2014xfa}
L.~Ma, G.~L.~Ma and Y.~G.~Ma,
``Anisotropic flow and flow fluctuations for Au + Au at $\sqrt{s_{\rm NN}}$ = 200 GeV in a multiphase transport model,''
Phys. Rev. C {\bf 89} (2014) 044907
[arXiv:1404.5935 [nucl-th]].

\bibitem{Niemi:2015qia}
H.~Niemi, K.~J.~Eskola and R.~Paatelainen,
``Event-by-event fluctuations in a perturbative QCD + saturation + hydrodynamics model: Determining QCD matter shear viscosity in ultrarelativistic heavy-ion collisions,''
Phys. Rev. C {\bf 93} (2016) 024907
[arXiv:1505.02677 [hep-ph]].

\bibitem{Giacalone:2016eyu}
G.~Giacalone, L.~Yan, J.~Noronha-Hostler and J.-Y.~Ollitrault,
``Skewness of elliptic flow fluctuations,''
Phys. Rev. C {\bf 95} (2017) 014913
[arXiv:1608.01823 [nucl-th]].

\bibitem{Bhalerao:2018anl}
R.~S.~Bhalerao, G.~Giacalone and J.-Y.~Ollitrault,
``Kurtosis of elliptic flow fluctuations,''
Phys. Rev. C {\bf 99} (2019) 014907
[arXiv:1811.00837 [nucl-th]].

\bibitem{Kapusta:2011gt}
J.~I.~Kapusta, B.~Muller and M.~Stephanov,
``Relativistic Theory of Hydrodynamic Fluctuations with Applications to Heavy Ion Collisions,''
Phys. Rev. C {\bf 85} (2012) 054906
[arXiv:1112.6405 [nucl-th]].

\bibitem{Nagle:2018nvi}
J.~L.~Nagle and W.~A.~Zajc,
``Small System Collectivity in Relativistic Hadron and Nuclear Collisions,''
Ann. Rev. Nucl. Part. Sci. {\bf 68} (2018) 211
[arXiv:1801.03477 [nucl-ex]].

\bibitem{Vogel:2007yq}
S.~Vogel, G.~Torrieri and M.~Bleicher,
``Elliptic flow fluctuations in heavy ion collisions at RHIC and the perfect fluid hypothesis,''
Phys. Rev. C {\bf 82} (2010) 024908
[arXiv:nucl-th/0703031 [nucl-th]].

\bibitem{Loizides:2014vua}
C.~Loizides, J.~Nagle and P.~Steinberg,
``Improved version of the PHOBOS Glauber Monte Carlo,''
SoftwareX {\bf 1-2} (2015) 13
[arXiv:1408.2549 [nucl-ex]].
	
\bibitem{Borghini:2018xum}
N.~Borghini, S.~Feld and N.~Kersting,
``Scaling behavior of anisotropic flow harmonics in the far from equilibrium regime,''
Eur. Phys. J. C \textbf{78} (2018) 832
[arXiv:1804.05729 [nucl-th]].

\bibitem{Kurkela:2018qeb}
A.~Kurkela, U.~A.~Wiedemann and B.~Wu,
``Opacity dependence of elliptic flow in kinetic theory,''
Eur. Phys. J. C {\bf 79} (2019) 759
[arXiv:1805.04081 [hep-ph]].

\bibitem{Gombeaud:2007ub}
C.~Gombeaud and J.-Y.~Ollitrault,
``Covariant transport theory approach to elliptic flow in relativistic heavy ion collision,''
Phys. Rev. C {\bf 77} (2008) 054904
[arXiv:nucl-th/0702075 [nucl-th]].
	
\bibitem{Bhalerao:2005mm}
R.~S.~Bhalerao, J.-P.~Blaizot, N.~Borghini and J.-Y. Ollitrault,
``Elliptic flow and incomplete equilibration at RHIC,''
Phys. Lett. B {\bf 627} (2005) 49
[arXiv:nucl-th/0508009 [nucl-th]].

\bibitem{Alver:2010dn}
B.~H.~Alver, C.~Gombeaud, M.~Luzum and J.-Y.~Ollitrault,
``Triangular flow in hydrodynamics and transport theory,''
Phys. Rev. C {\bf 82} (2010) 034913
[arXiv:1007.5469 [nucl-th]].

\bibitem{Kurkela:2020wwb}
A.~Kurkela, S.~F.~Taghavi, U.~A.~Wiedemann and B.~Wu,
``Hydrodynamization in systems with detailed transverse profiles,''
Phys. Lett. B {\bf 811} (2020) 135901
[arXiv:2007.06851 [hep-ph]].

\bibitem{Jackknife}
J.~Shao and D.~Tu, {\em The Jackknife and Bootstrap}.
(Springer, New York, 1995).

\bibitem{Borghini:2005kd}
N.~Borghini and J.-Y.~Ollitrault,
``Momentum spectra, anisotropic flow, and ideal fluids,''
Phys.\ Lett.\ B {\bf 642} (2006) 227  
[nucl-th/0506045].

\bibitem{Teaney:2012ke}
D.~Teaney and L.~Yan,
``Nonlinearities in the harmonic spectrum of heavy ion collisions with ideal and viscous hydrodynamics,''
Phys.\ Rev.\ C {\bf 86} (2012) 044908  
[arXiv:1206.1905 [nucl-th]].

\end{thebibliography}
\end{document}